\documentclass[a4paper,fleqn,usenatbib,useAMS]{mnras}

\usepackage{graphicx}	
\usepackage{amsmath}	
\usepackage{amssymb}	
\usepackage{multicol}        
\usepackage{bm}		
\usepackage{pdflscape}	

\newcommand{\cps}{\,counts\,sec$^{-1}$}

\usepackage[T1]{fontenc}
\usepackage{ae,aecompl}
\usepackage{graphicx}
\usepackage{txfonts}
\usepackage{booktabs}
\usepackage{threeparttable}
\usepackage{subcaption}

\title[Nature of MAXI J0637-430]{Revealing the nature of the transient source MAXI J0637-430 through spectro-temporal analysis}
 
\author[B. E. Baby et al.]{Blessy E. Baby$^{1}$\thanks{Contact e-mail: \href{mailto:blessy.elizabeth65@gmail.com}{blessy.elizabeth65@gmail.com}}, G. R. Bhuvana$^{2}$, D. Radhika$^{2}$, Tilak Katoch$^{3}$, Samir Mandal$^{4}$ and \newauthor{Anuj Nandi$^{5}$}
\\
$^{1}$Department of Physics, University of Calicut, Malappuram 673635, Kerala, India.\\
$^{2}$Department of Physics, Dayananda Sagar University, Hosur Main Road, Bangalore 560068, Karnataka, India.\\
$^{3}$Tata Institute of Fundamental Research, Homi Bhabha Road, Mumbai 400005, Maharashtra, India.\\
$^{4}$Department of Earth and Space Sciences, Indian Institute of Space Science and Technology, Thiruvananthapuram 695547, Kerala, India.\\
$^{5}$Space Astronomy Group, ISITE campus, U. R. Rao Satellite Centre, Karthik Nagar, Bangalore 560037, Karnataka, India.\\
}

\date{Last updated 2015 May 22; in original form 2013 September 5}

\pubyear{2021}

\begin{document}
\label{firstpage}
\pagerange{\pageref{firstpage}--\pageref{lastpage}}
\maketitle

\begin{abstract}
 We study the spectral and temporal properties of MAXI J0637-430 during its 2019-2020 outburst using \textit{NICER}, \textit{AstroSat} and \textit{Swift-XRT} data. The source was in a disc dominant state within a day of its detection and traces out a `c' shaped profile in the HID, similar to the `mini'-outbursts of the recurrent BHB 4U 1630-472. Energy spectrum is obtained in the $0.5-10$ keV band with \textit{NICER} and \textit{Swift-XRT}, and $0.5-25$ keV with \textit{AstroSat}. The spectra can be modelled using a multicolour disc emission (\textit{diskbb}) convolved with a thermal Comptonisation component (\textit{thcomp}). The disc temperature decreases from 0.6 keV to 0.1 keV during the decay with a corresponding decrease in photon index ($\Gamma$) from 4.6 to 1.8. The fraction of Compton scattered photons ($f_{cov}$) remains $<$ 0.3 during the decay upto mid-January 2020 and gradually increases to 1 as the source reaches hard state. Power Density Spectra (PDS) generated in the 0.01-100 Hz range display no Quasi-periodic Oscillations (QPOs) although band-limited noise (BLN) is seen towards the end of January 2020. During \textit{AstroSat} observations, $\Gamma$ lies in the range $2.3-2.6$ and rms increases from 11 to 20\%, suggesting that the source was in an intermediate state till 21 November 2019. Spectral fitting with the relativistic disc model (\textit{kerrbb}), in conjunction with the soft-hard transition luminosity, favour a black hole with mass $3-19$ $M_{\sun}$ with retrograde spin at a distance $<15$ kpc. Finally, we discuss the possible implications of our findings. 
\end{abstract}

\begin{keywords}
   accretion, accretion discs -- black hole physics -- X-rays:binaries -- stars:black holes -- stars:individual:MAXI J0637-430
\end{keywords}




\section{Introduction}

The study of X-ray binaries (XRBs) is done in pursuit of a better understanding of complex physical phenomena in strong gravitational fields. XRBs are divided into High Mass XRBs (HMXB) and Low Mass XRBs (LMXB) based on the mass of the companion star \citep[][and references therein]{Bildsten1997,Corral2016}. While some sources remain consistently bright (persistent sources), others display recurring episodes of increase in brightness followed by a decay into quiescence, called an outburst (transients). The duration and recurrence period of these outbursts vary from days to months for different sources and sometimes different outbursts in the same source itself \citep[][and references therein]{Homan2001,Belloni2005,Lewin2006,Nandi2012,Baby2020}. 

Black Hole Binaries (BHBs) are XRBs with a black hole as the compact object. Modelling the energy spectra of these sources helps in understanding the physical processes which power the outbursts. The spectra can generally be modelled using a multicolour black body component arising from the accretion disc \citep{Shakura1973} and a powerlaw component corresponding to the  Comptonisation of the thermal disc photons by the high energy electrons in the corona \citep{Sunyaev1980,Chakrabarti1995,Narayan1996}. The components' relative contribution varies as the source passes through different stages in its outburst leading to the classification of spectral states. The outbursts are classified into Low/Hard State (LHS), Hard Intermediate State (HIMS), Soft Intermediate State (SIMS) and High/Soft State (HSS) \cite[][and references therein]{Homan2001,Fender2004,Homan2005,Belloni2005,Remillard2006,Belloni2010,Nandi2012,Radhika2014,Sreehari2019,Baby2020}. The source spectra evolve from being powerlaw dominant in the LHS to disc dominant in the HSS. Apart from the spectral changes, temporal evolution is also seen in different states, which is studied using the Fourier transform of the lightcurve. The PDS changes from a weak powerlaw with featureless noise (in HSS) to band-limited noise which can either evolve to or is accompanied by narrow, peaked Lorentzian features, termed QPOs (in LHS and intermediate states). The evolution of the source through an outburst can be tracked in the Hardness-Intensity Diagram (HID) which is a plot between the ratio of flux in the hard to soft energy band and the total flux. These HIDs usually follow a `q'-shaped profile in sources where the canonical state transitions are observed  \citep[][and references therein]{Homan2001,Fender2004,Homan2005,Belloni2005,Remillard2006,Nandi2012,Radhika2014}. Study of the spectral and temporal properties of the source in conjunction with HID not only helps in understanding the nature of the source as a single entity but also aids in contextualizing its properties with reference to other `standard' sources. Such a categorization is particularly helpful in the case of newly discovered sources. In the present work, we aim to study the nature and accretion flow geometry of the newly discovered X-ray transient MAXI J0637-430.

  MAXI J0637-430 was discovered by the \textit{Monitor of All-Sky X-ray Image (MAXI)} alert system on 2 November 2019, at 06:37 UT \citep{Negoro2019}. Following the detection, \textit{Neil Gehrels SWIFT observatory} observed the source on 3 November 2019, and obtained the position of the source as RA $06^{h}$ $36^{m}$ $23.59^{s}$ and Dec $-42^{d}$ $52^{m}$ $04.1^{s}$ \citep{Kennea2019}. Optical observations followed, which show a double-peaked H$\alpha$ and He-II 4686 emission and weak emission from H$\beta$, H$\gamma$ and He-I 5875 and 6678 in the spectrum \citep{Strader2019}. An optical companion was detected which is observed to have a blue spectrum with a slope of -3.22, and reddening factor of E(B-V) = 0.064 with average source magnitude of g=16.1 \citep{Li2019}. An IR counterpart was also detected close to the optical counterpart \citep{Murata2019}. Radio detection was reported using \textit{Australia Telescope Compact Array (ATCA)} on 6 November 2019 \citep{Russell2019}. In X-rays, the source seems to be a soft XRB with disc temperature of $\sim$ 0.68 keV and powerlaw photon index of $\sim$ 2.4, confirming it to be a black hole XRB \citep{Tomsick2019}. Positive residuals at Fe K region and above 20 keV are reported using \textit{Nuclear Spectroscopic Telescope Array (NuSTAR)} data, although timing analysis did not show any substantial variability \citep{Tomsick2019}. Preliminary spectral analysis of \textit{AstroSat} observation \citep{Thomas2019} of the source on 8 November 2019 resulted in photon index of $\sim$ 1.98, which is harder relative to \textit{NuSTAR} observation, although the spectrum remains predominantly soft. Transition to a hard state was later reported on 14 January 2020 using \textit{Neutron Star Interior Composition Explorer (NICER)} observations. An optical re-brightening \citep{Baglio2020} was also reported around the transition to the hard state. The optical flux peaked on 4 February 2020 and declined to quiescence by the end of April \citep{Johar2020}. Recently, \cite{Tetarenko2021} performed multi-wavelength spectral analysis of X-ray, UV and optical data from which an orbital period of $2-4$ hours for the system was obtained assuming standard parameters for LMXBs \citep{Tetarenko2016}. The multi-wavelength spectrum was modelled with an irradiated disc and a powerlaw component. \cite{Jana2021} performed spectro-temporal analysis using \textit{NICER} data and classified the outburst into SIMS, HSS, HIMS and LHS. The mass of the compact object, assuming a non-rotating black hole at distance $<$ 10 kpc, is estimated to be in the range of $5-12$ $M_{\sun}$. \cite{Lazar2021} state that a two component model (thermal disc and Comptonisation component) is not enough to explain the soft state spectrum of the source when the broadband spectrum from \textit{NuSTAR} is considered. They suggest that emission from the plunging region or reprocessing of the returning disc radiation are equally plausible physical scenarios based on spectral modelling. 
 
In this work, we study the evolutionary track of the source during the outburst by performing detailed spectro-temporal analysis on \textit{NICER} \citep{Gendreau2016} and \textit{X-ray Telescope onboard Neil Gehrels Swift Observatory (Swift-XRT)} \citep{Burrows2005} data, aided by the existence of broadband spectra from \textit{AstroSat} \citep{Agrawal2001,Agrawal2006,Singh2014}. More robust classification of the outburst into different states is attempted. We also use physical models and transition luminosity to remark on the nature of the source. 

In Section 2, details on observation and data reduction are mentioned. In Section 3, spectral and temporal analysis techniques are discussed along with the models used. We comment on the evolution of the spectro-temporal properties of the source in Section 4. Finally, we discuss the nature of the source and present our conclusions in Section 5.  

\section{Observations and Data Reduction}

We obtained data from three different instruments - \textit{NICER, AstroSat} and \textit{Swift-XRT} to study the evolution of the source during the outburst. The details of observations chosen are given in Tables \ref{tab:pheno_fit_nicer}, \ref{tab:pheno_fit_astro} and \ref{tab:pheno_fit_swift}. 
   \begin{figure}
   \centering
   \includegraphics[scale=0.30]{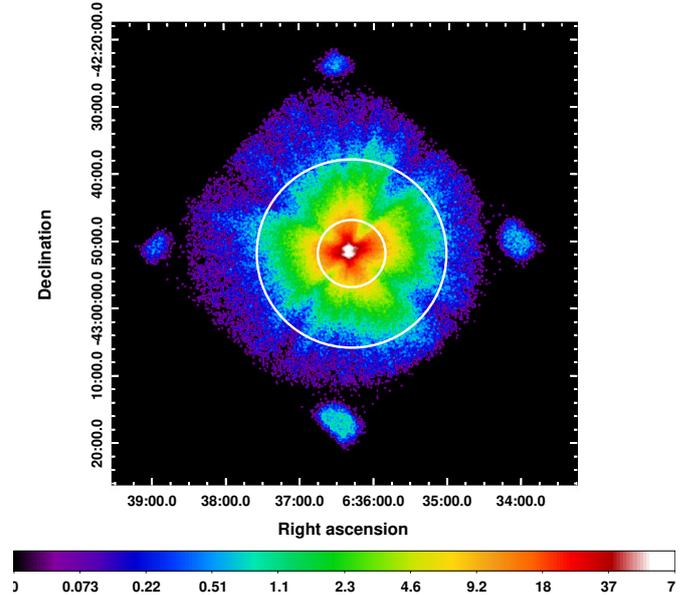}
   \caption{\textit{SXT} image of the source observed on 21 November 2019. Annular source region marked with inner and outer radius of $5'$ and $14'$ respectively is shown. The colour bar shows the variation in flux.}
              \label{fig:sxt_img}%
    \end{figure}
\subsection{\textit{NICER}}
\textit{NICER} observed MAXI J0637-430 at regular intervals since 3 November 2019 till the end of April 2020. We obtained all available data for the source between 3 November 2019 and 30 January 2020. For observations after 3 February 2020 to March 2020, where the source count rate is below 10 \cps and variation in hardness ratios and spectral parameters is minimal, we considered only those observations where the exposure time was $>$ 1.5 ks. Details of the observations and the count rates are given in Table \ref{tab:pheno_fit_nicer}. We processed these observations using \texttt{HEASOFT} version 6.28 and \texttt{NICERDAS} version 7.0 to obtain the clean event file\footnote{\tiny{\url{https://heasarc.gsfc.nasa.gov/docs/nicer/nicer_analysis.html}}}. The latest CALDB files (20200722) were used. The event file was screened considering the standard criteria i.e., a pointing offset $<54''$,  dark earth limb $>30^{\circ}$, bright Earth limb $>40^{\circ}$ and outside the South Atlantic Anomaly (SAA) region. Data from detectors 14 and 34 was removed using \texttt{fselect} command as they are reported to show episodes of increased electronic noise. Time intervals showing strong background flare-ups where the average count rate in the $13-15$ keV band is greater than 1 \cps were excluded to generate the final Good Time Intervals (GTIs). The background spectra were calculated using the 3C50\_RGv5 model provided by \textit{NICER} team \citep{Remillard2021}. The latest version of Redistribution Matrix File (RMF) provided by the team was used. Ancillary Response Files (ARF) for 50 detectors were co-added to generate the final ARF file. The spectra were rebinned to have a minimum of 25 counts per bin.

To create the lightcurve and obtain hardness ratios, we first produced 1 s binned lightcurves in the $0.5-10$, $0.5-2$ and $2-10$ keV bands using \texttt{XSELECT}. The Hardness ratio is defined as the ratio of counts in the $2-10$ keV band to the counts in $0.5-2$ keV band. The lightcurves in the $0.3-12$ keV range with a binsize of 5 ms were used to generate the PDS.  

   \begin{figure}
   \centering
   \includegraphics[scale=0.5]{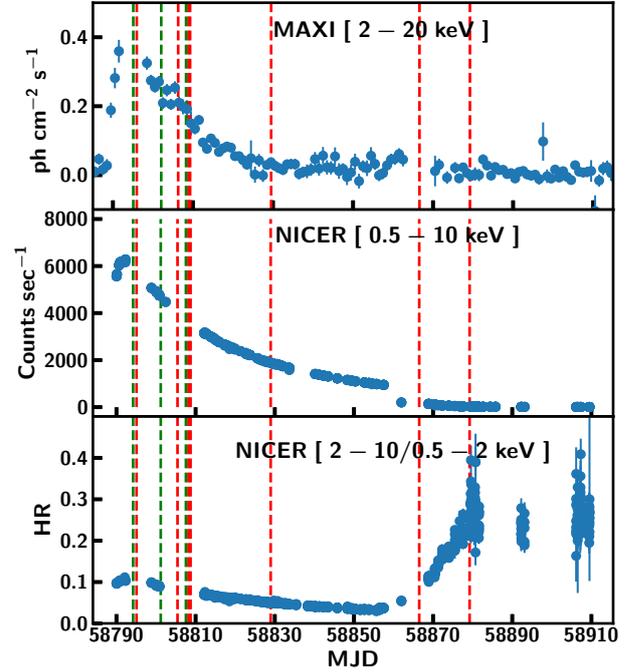}
      \caption{\textit{MAXI} lightcurve in the $2-20$ keV band is plotted in the top panel. \textit{NICER} lightcurve in the $0.5-10$ keV band is shown in the middle panel. The bottom panel shows the hardness ratios obtained using \textit{NICER} data. HR is defined as the ratio of counts in $2-10$ keV to counts in $0.5-2$ keV. \textit{NICER} lightcurve is binned by 100 seconds for clarity. \textit{AstroSat} and \textit{Swift-XRT} observations are marked by green and red dashed vertical lines.}
         \label{fig:maxi_fig}
   \end{figure}

\begin{table*}\small
  \caption[]{Details of \textit{NICER} observations chosen for analysis are provided in the first 5 columns. Spectral parameters fitted with model\\
  		 \textit{TBabs (thcomp $\otimes$ diskbb)} are presented with 90\% confidence.}
     \label{tab:pheno_fit_nicer}
\begin{threeparttable}      
\begin{tabular}{c@{\hspace{2pt}}cc@{\hspace{2pt}}c@{\hspace{2pt}}cc@{\hspace{2pt}}cccc@{\hspace{2pt}}cc}
\hline
\smallskip
ObsId & Date & MJD & Exp. time & Rate & $n_{H}$ & $\Gamma$ & $f_{cov}^{a}$ & $T_{in}$ & N$_{diskbb}$ & $F_{0.5-10\;keV}$ & $\chi^{2}/dof$ \\
 & & & \scriptsize{(ks)} & \scriptsize{\cps} & \scriptsize{$(\times 10^{20}$ cm$^{-2})$} & & & \scriptsize{(keV)} & & \scriptsize{$(\times\;10^{-10}$ erg cm$^{-2}$ s$^{-1}$)} & \\
\hline
2200950101 & 2019-11-03 & 58790.89 & 1.1 & 5607 & $2.3_{-0.1}^{+0.1}$ & $4.60_{-0.02}^{+0.10}$ & $0.20\pm0.10$ & $0.58_{-0.002}^{+0.001}$ & $3606_{-62}^{+19}$ & $72.00 \pm {0.20}$ & 457/556 \\
2200950102 & 2019-11-04 & 58791.02 & 4.5 & 6140 & $2.5_{-0.2}^{+0.2}$ & $4.50_{-0.20}^{+0.30}$ & $0.32_{-0.06}^{+0.08}$ & $0.59_{-0.003}^{+0.003}$ & $3731_{-87}^{+19}$ & $79.40 \pm {0.20}$ & 576/702 \\
2200950103 & 2019-11-05 & 58792.05 & 0.6 & 6226 & $2.6_{-0.2}^{+0.2}$ & $4.30_{-0.10}^{+0.10}$ & $0.25\pm0.07$ & $0.59_{-0.002}^{+0.002}$ & $3673_{-62}^{+19}$ & $81.50 \pm {0.40}$ & 540/534 \\
2200950104 & 2019-11-06 & 58793.15 & 0.3 & 6271 & $2.8_{-0.3}^{+0.3}$ & $4.00_{-0.10}^{+0.10}$ & $0.25\pm0.08$ & $0.59_{-0.003}^{+0.003}$ & $3826_{-75}^{+19}$ & $82.70 \pm {0.30}$ & 428/501 \\
2200950107 & 2019-11-12 & 58799.70 & 0.6 & 5068 & $2.0_{-0.1}^{+0.2}$ & $2.90_{-0.03}^{+0.03}$ & $0.25\pm0.03$ & $0.55_{-0.002}^{+0.002}$ & $3784_{-72}^{+19}$ & $65.70 \pm {0.10}$ & 562/573 \\
2200950108 & 2019-11-13 & 58800.22 & 1.2 & 4892 & $2.1_{-0.2}^{+0.2}$ & $3.20_{-0.03}^{+0.04}$ & $0.15\pm0.02$ & $0.56_{-0.002}^{+0.002}$ & $3594_{-51}^{+19}$ & $61.50 \pm {0.20}$ & 544/585 \\
2200950109 & 2019-11-14 & 58801.05 & 1.1 & 4759 & $2.2_{-0.2}^{+0.2}$ & $3.76_{-0.05}^{+0.05}$ & $0.25\pm0.04$ & $0.55_{-0.002}^{+0.002}$ & $3746_{-61}^{+19}$ & $60.90 \pm {0.30}$ & 526/589 \\
2200950112a & 2019-11-26 & 58813.03 & 3.2 & 3140 & $2.1_{-0.2}^{+0.1}$ & $4.09_{-0.04}^{+0.04}$ & $0.25\pm0.03$ & $0.49_{-0.001}^{+0.001}$ & $3722_{-52}^{+19}$ & $35.90 \pm {0.20}$ & 639/577 \\
2200950112b & 2019-11-26 & 58813.48 & 2.5 & 3111 & $2.0_{-0.2}^{+0.2}$ & $4.00_{-0.20}^{+0.20}$ & $0.31_{-0.06}^{+0.07}$ & $0.49_{-0.004}^{+0.004}$ & $3824_{-127}^{+19}$ & $35.60 \pm {0.30}$ & 562/570 \\
2200950113 & 2019-11-27 & 58814.77 & 1.2 & 2988 & $2.1_{-0.2}^{+0.2}$ & $3.70_{-0.30}^{+0.30}$ & $0.26_{-0.03}^{+0.09}$ & $0.48_{-0.005}^{+0.004}$ & $3892_{-150}^{+19}$ & $34.20 \pm {0.20}$ & 504/505 \\
2200950114 & 2019-11-28 & 58815.22 & 1.9 & 2913 & $2.1_{-0.4}^{+0.2}$ & $4.20_{-0.30}^{+0.30}$ & $0.40_{-0.10}^{+0.10}$ & $0.47_{-0.005}^{+0.004}$ & $3980_{-157}^{+19}$ & $33.00 \pm {0.20}$ & 575/537 \\
2200950115 & 2019-11-29 & 58816.26 & 2.0 & 2789 & $2.0_{-0.2}^{+0.2}$ & $4.45_{-0.05}^{+0.06}$ & $0.35\pm0.05$ & $0.47_{-0.002}^{+0.002}$ & $3825_{-34}^{+19}$ & $31.30 \pm {0.20}$ & 520/516 \\
2200950116 & 2019-11-30 & 58817.60 & 1.1 & 2725 & $2.2_{-0.2}^{+0.2}$ & $4.14_{-0.07}^{+0.07}$ & $0.25\pm0.05$ & $0.47_{-0.002}^{+0.002}$ & $3826_{-73}^{+19}$ & $30.70 \pm {0.30}$ & 520/464 \\
2200950117 & 2019-12-01 & 58818.19 & 1.8 & 2638 & $1.8_{-0.2}^{+0.1}$ & $4.19_{-0.06}^{+0.06}$ & $0.25\pm0.04$ & $0.47_{-0.002}^{+0.002}$ & $3658_{-32}^{+19}$ & $29.40 \pm {0.10}$ & 561/494 \\
2200950118 & 2019-12-02 & 58819.42 & 0.7 & 2570 & $2.2_{-0.2}^{+0.2}$ & $3.96_{-0.07}^{+0.07}$ & $0.25\pm0.06$ & $0.46_{-0.001}^{+0.002}$ & $3904_{-88}^{+19}$ & $28.60 \pm {0.20}$ & 460/435 \\
2200950119 & 2019-12-03 & 58820.19 & 2.9 & 2482 & $1.8_{-0.2}^{+0.2}$ & $4.00_{-0.30}^{+0.20}$ & $0.29_{-0.06}^{+0.07}$ & $0.46_{-0.004}^{+0.004}$ & $3772_{-132}^{+19}$ & $27.20 \pm {0.10}$ & 630/547 \\
2200950120 & 2019-12-05 & 58822.19 & 0.6 & 2379 & $1.9_{-0.2}^{+0.2}$ & $3.72_{-0.06}^{+0.06}$ & $0.25\pm0.05$ & $0.45_{-0.002}^{+0.002}$ & $3896_{-99}^{+19}$ & $26.40 \pm {0.10}$ & 452/420 \\
2200950121 & 2019-12-06 & 58823.50 & 1.9 & 2277 & $1.8_{-0.2}^{+0.2}$ & $3.75_{-0.04}^{+0.04}$ & $0.25\pm0.03$ & $0.45_{-0.002}^{+0.002}$ & $3856_{-73}^{+19}$ & $24.80 \pm {0.10}$ & 602/514 \\
2200950122 & 2019-12-07 & 58824.47 & 1.8 & 2217 & $2.1_{-0.3}^{+0.2}$ & $4.00_{-0.20}^{+0.30}$ & $0.36_{-0.08}^{+0.10}$ & $0.44_{-0.005}^{+0.005}$ & $4152_{-188}^{+19}$ & $24.10 \pm {0.10}$ & 513/508 \\
2200950123 & 2019-12-08 & 58825.80 & 1.2 & 2124 & $1.9_{-0.2}^{+0.2}$ & $3.92_{-0.05}^{+0.05}$ & $0.30\pm0.05$ & $0.43_{-0.002}^{+0.002}$ & $3987_{-89}^{+19}$ & $23.00 \pm {0.20}$ & 474/459 \\
2200950124 & 2019-12-09 & 58826.53 & 2.1 & 2071 & $1.7_{-0.2}^{+0.2}$ & $4.08_{-0.03}^{+0.04}$ & $0.30\pm0.04$ & $0.43_{-0.002}^{+0.003}$ & $3886_{-76}^{+19}$ & $22.30 \pm {0.10}$ & 592/505 \\
2200950125 & 2019-12-10 & 58827.57 & 0.9 & 1994 & $2.0_{-0.2}^{+0.2}$ & $4.50_{-0.10}^{+0.10}$ & $0.30\pm0.08$ & $0.43_{-0.002}^{+0.002}$ & $3881_{-95}^{+19}$ & $21.20 \pm {0.20}$ & 587/413 \\
2200950126 & 2019-12-11 & 58828.34 & 1.4 & 1944 & $1.8_{-0.1}^{+0.2}$ & $4.38_{-0.06}^{+0.07}$ & $0.30\pm0.06$ & $0.43_{-0.002}^{+0.002}$ & $3840_{-84}^{+19}$ & $20.60 \pm {0.10}$ & 524/441 \\
2200950127 & 2019-12-12 & 58829.11 & 1.2 & 1920 & $1.9_{-0.2}^{+0.2}$ & $4.24_{-0.06}^{+0.07}$ & $0.30\pm0.06$ & $0.42_{-0.002}^{+0.002}$ & $3969_{-92}^{+19}$ & $20.30 \pm {0.20}$ & 493/433 \\
2200950128 & 2019-12-13 & 58830.08 & 2.1 & 1852 & $1.8_{-0.2}^{+0.2}$ & $4.26_{-0.05}^{+0.05}$ & $0.30\pm0.05$ & $0.42_{-0.002}^{+0.002}$ & $3855_{-78}^{+19}$ & $19.50 \pm {0.10}$ & 532/468 \\
2200950129 & 2019-12-14 & 58831.04 & 3.0 & 1822 & $1.9_{-0.2}^{+0.2}$ & $4.15_{-0.04}^{+0.04}$ & $0.30\pm0.04$ & $0.42_{-0.002}^{+0.002}$ & $3923_{-72}^{+19}$ & $19.10 \pm {0.20}$ & 596/502 \\
2200950130 & 2019-12-15 & 58832.32 & 2.0 & 1770 & $2.1_{-0.2}^{+0.1}$ & $4.04_{-0.04}^{+0.05}$ & $0.30\pm0.04$ & $0.41_{-0.002}^{+0.002}$ & $4078_{-83}^{+19}$ & $18.50 \pm {0.20}$ & 547/471 \\
2200950131 & 2019-12-16 & 58833.35 & 1.0 & 1722 & $2.2_{-0.3}^{+0.2}$ & $4.50_{-0.10}^{+0.10}$ & $0.30\pm0.07$ & $0.41_{-0.002}^{+0.003}$ & $3994_{-99}^{+19}$ & $18.00 \pm {0.10}$ & 399/391 \\
2200950132 & 2019-12-17 & 58834.39 & 1.4 & 1673 & $1.7_{-0.1}^{+0.2}$ & $4.47_{-0.07}^{+0.06}$ & $0.30\pm0.07$ & $0.41_{-0.002}^{+0.002}$ & $3818_{-91}^{+19}$ & $17.30 \pm {0.20}$ & 462/419 \\
2200950133 & 2019-12-18 & 58835.99 & 0.8 & 1605 & $2.1_{-0.3}^{+0.4}$ & $4.01_{-0.07}^{+0.07}$ & $0.30\pm0.06$ & $0.40_{-0.002}^{+0.002}$ & $4204_{-123}^{+19}$ & $16.70 \pm {0.30}$ & 423/389 \\
2200950135 & 2019-12-22 & 58839.22 & 0.8 & 1470 & $2.4_{-0.4}^{+0.4}$ & $4.40_{-0.50}^{+0.50}$ & $0.40_{-0.20}^{+0.20}$ & $0.38_{-0.009}^{+0.007}$ & $4634_{-369}^{+19}$ & $14.90 \pm {0.20}$ & 429/379 \\
2200950136 & 2019-12-23 & 58840.06 & 0.5 & 1446 & $2.7_{-0.3}^{+0.3}$ & $4.10_{-0.10}^{+0.10}$ & $0.30\pm0.09$ & $0.38_{-0.003}^{+0.003}$ & $4596_{-153}^{+19}$ & $14.80 \pm {0.20}$ & 432/336 \\
2200950137 & 2019-12-24 & 58841.09 & 0.8 & 1405 & $2.0_{-0.3}^{+0.3}$ & $4.19_{-0.09}^{+0.09}$ & $0.30\pm0.08$ & $0.38_{-0.002}^{+0.002}$ & $4213_{-127}^{+19}$ & $14.20 \pm {0.10}$ & 495/386 \\
2200950138 & 2019-12-25 & 58842.13 & 1.0 & 1371 & $2.1_{-0.3}^{+0.3}$ & $4.12_{-0.07}^{+0.07}$ & $0.30\pm0.07$ & $0.38_{-0.002}^{+0.002}$ & $4220_{-124}^{+19}$ & $13.90 \pm {0.10}$ & 472/377 \\
2200950139 & 2019-12-26 & 58843.16 & 1.0 & 1344 & $1.6_{-0.2}^{+0.2}$ & $4.02_{-0.06}^{+0.07}$ & $0.30\pm0.06$ & $0.38_{-0.002}^{+0.002}$ & $4320_{-140}^{+19}$ & $13.70 \pm {0.10}$ & 484/381 \\
2200950140 & 2019-12-27 & 58844.32 & 0.9 & 1307 & $2.0_{-0.4}^{+0.4}$ & $3.40_{-0.30}^{+0.30}$ & $0.20_{-0.06}^{+0.09}$ & $0.38_{-0.006}^{+0.005}$ & $4289_{-249}^{+19}$ & $13.00 \pm {0.10}$ & 459/385 \\
2200950142 & 2019-12-29 & 58846.66 & 1.1 & 1224 & $2.0_{-0.3}^{+0.3}$ & $4.00_{-0.06}^{+0.06}$ & $0.30\pm0.06$ & $0.37_{-0.003}^{+0.002}$ & $4324_{-132}^{+19}$ & $12.10 \pm {0.20}$ & 486/372 \\
2200950144 & 2020-01-01 & 58849.22 & 1.7 & 1150 & $2.0_{-0.2}^{+0.3}$ & $4.12_{-0.05}^{+0.05}$ & $0.30\pm0.10$ & $0.36_{-0.002}^{+0.002}$ & $4392_{-126}^{+19}$ & $11.40 \pm {0.10}$ & 518/421 \\
2200950145 & 2020-01-02 & 58850.26 & 1.9 & 1123 & $2.0_{-0.2}^{+0.2}$ & $3.90_{-0.04}^{+0.04}$ & $0.30\pm0.04$ & $0.36_{-0.002}^{+0.002}$ & $4402_{-120}^{+19}$ & $11.00 \pm {0.10}$ & 486/409 \\
2200950146 & 2020-01-03 & 58851.23 & 1.1 & 1098 & $2.5_{-0.3}^{+0.3}$ & $4.15_{-0.07}^{+0.08}$ & $0.30\pm0.07$ & $0.36_{-0.002}^{+0.002}$ & $4543_{-138}^{+19}$ & $10.80 \pm {0.10}$ & 497/360 \\
2200950147 & 2020-01-04 & 58852.46 & 1.3 & 1061 & $1.8_{-0.4}^{+0.3}$ & $4.28_{-0.07}^{+0.07}$ & $0.30\pm0.06$ & $0.36_{-0.002}^{+0.002}$ & $4143_{-123}^{+19}$ & $10.30 \pm {0.30}$ & 448/359 \\
2200950148 & 2020-01-05 & 58853.49 & 1.2 & 1049 & $2.1_{-0.3}^{+0.3}$ & $4.05_{-0.06}^{+0.06}$ & $0.30\pm0.06$ & $0.36_{-0.002}^{+0.002}$ & $4440_{-142}^{+19}$ & $10.20 \pm {0.20}$ & 398/360 \\
2200950149 & 2020-01-06 & 58854.65 & 1.1 & 1026 & $2.5_{-0.4}^{+0.3}$ & $4.20_{-0.30}^{+0.40}$ & $0.50_{-0.10}^{+0.20}$ & $0.34_{-0.008}^{+0.006}$ & $5024_{-576}^{+19}$ & $10.00 \pm {0.20}$ & 432/359 \\
2200950151 & 2020-01-08 & 58856.52 & 2.7 & 979 & $2.6_{-0.4}^{+0.4}$ & $4.40_{-0.20}^{+0.20}$ & $0.50_{-0.10}^{+0.10}$ & $0.34_{-0.006}^{+0.005}$ & $5241_{-344}^{+19}$ & $9.50 \pm {0.10}$ & 434/424 \\
2200950152 & 2020-01-09 & 58857.37 & 1.2 & 968 & $2.1_{-0.4}^{+0.4}$ & $3.30_{-0.20}^{+0.20}$ & $0.24_{-0.06}^{+0.08}$ & $0.34_{-0.005}^{+0.005}$ & $4585_{-289}^{+19}$ & $9.30 \pm {0.10}$ & 442/389 \\
2200950153 & 2020-01-10 & 58858.34 & 1.0 & 945 & $2.6_{-0.4}^{+0.5}$ & $3.20_{-0.20}^{+0.20}$ & $0.34_{-0.06}^{+0.08}$ & $0.33_{-0.006}^{+0.006}$ & $5383_{-422}^{+19}$ & $9.40 \pm {0.10}$ & 446/397 \\
2200950154 & 2020-01-14 & 58862.85 & 0.3 & 197 & $2.5^{f}$ & $2.21_{-0.09}^{+0.09}$ & $0.29_{-0.04}^{+0.04}$ & $0.19_{-0.005}^{+0.005}$ & $10636_{-1073}^{+19}$ & $1.96 \pm {0.03}$ & 221/219 \\
2200950155 & 2020-01-21 & 58869.71 & 1.1 & 134 & $2.5^{f}$ & $1.89_{-0.03}^{+0.03}$ & $0.27_{-0.01}^{+0.01}$ & $0.13_{-0.003}^{+0.003}$ & $46265_{-5144}^{+19}$ & $1.54 \pm {0.02}$ & 355/383 \\
2200950156 & 2020-01-23 & 58871.19 & 1.1 & 109 & $2.5^{f}$ & $1.86_{-0.03}^{+0.03}$ & $0.29_{-0.01}^{+0.01}$ & $0.12_{-0.003}^{+0.003}$ & $47390_{-6368}^{+19}$ & $1.34 \pm {0.02}$ & 348/375 \\
2200950157 & 2020-01-24 & 58872.03 & 1.4 & 82 & $2.5^{f}$ & $1.84_{-0.03}^{+0.02}$ & $0.34_{-0.02}^{+0.02}$ & $0.11_{-0.004}^{+0.004}$ & $49297_{-9675}^{+19}$ & $1.12 \pm {0.02}$ & 376/388 \\
2200950158 & 2020-01-25 & 58873.19 & 0.8 & 64 & $2.5^{f}$ & $1.78_{-0.04}^{+0.04}$ & $0.39_{-0.02}^{+0.03}$ & $0.12_{-0.006}^{+0.006}$ & $27830_{-6985}^{+19}$ & $0.95 \pm {0.02}$ & 260/298 \\
2200950159 & 2020-01-26 & 58874.81 & 2.5 & 58 & $2.5^{f}$ & $1.82_{-0.02}^{+0.02}$ & $0.46_{-0.02}^{+0.02}$ & $0.11_{-0.005}^{+0.005}$ & $33195_{-7506}^{+19}$ & $0.87 \pm {0.01}$ & 437/449 \\
2200950160 & 2020-01-28 & 58876.36 & 1.2 & 44 & $2.5^{f}$ & $1.82_{-0.03}^{+0.03}$ & $0.58_{-0.05}^{+0.04}$ & $0.11_{-0.010}^{+0.010}$ & $19041_{-6906}^{+19}$ & $0.71 \pm {0.01}$ & 334/333 \\
2200950161 & 2020-01-29 & 58877.58 & 1.2 & 30 & $2.5^{f}$ & $1.72_{-0.04}^{+0.03}$ & $0.73_{-0.06}^{+0.06}$ & $0.12_{-0.010}^{+0.020}$ & $9619_{-3691}^{+19}$ & $0.59 \pm {0.01}$ & 308/299 \\
2200950162 & 2020-01-30 & 58878.42 & 1.3 & 26 & $2.5^{f}$ & $1.78_{-0.04}^{+0.03}$ & $0.85_{-0.08}^{+0.08}$ & $0.12_{-0.010}^{+0.020}$ & $8073_{-4317}^{+19}$ & $0.50 \pm {0.02}$ & 326/298 \\
2200950164 & 2020-02-01 & 58880.23 & 5.0 & 20 & $2.5^{f}$ & $1.69_{-0.02}^{+0.02}$ & $0.92_{-0.04}^{+0.04}$ & $0.14_{-0.020}^{+0.020}$ & $2852_{-1112}^{+19}$ & $0.44 \pm {0.02}$ & 535/525 \\
2200950165 & 2020-02-02 & 58881.07 & 3.1 & 17 & $2.5^{f}$ & $1.76_{-0.02}^{+0.01}$ & $1.00\pm0.10$ & $0.10^{f}$ & $8974_{-135}^{+19}$ & $0.36 \pm {0.04}$ & 369/397 \\
2200950166 & 2020-02-03 & 58882.43 & 1.6 & 15 & $2.5^{f}$ & $1.77_{-0.03}^{+0.02}$ & $1.00\pm0.10$ & $0.10^{f}$ & $7995_{-172}^{+19}$ & $0.32 \pm {0.02}$ & 306/282 \\
2200950176 & 2020-02-14 & 58893.13 & 1.6 & 8 & $2.5^{f}$ & $1.72_{-0.03}^{+0.02}$ & $1.00\pm0.20$ & $0.10^{f}$ & $4287_{-126}^{+19}$ & $0.18 \pm {0.02}$ & 187/220 \\
2200950188 & 2020-02-28 & 58907.04 & 1.9 & 5 & $2.5^{f}$ & $1.70_{-0.03}^{+0.04}$ & $1.00 \pm 0.20$ & $0.10^{f}$ & $2424_{-76}^{+19}$ & $0.11 \pm {0.02}$ & 200/199 \\
2200950189 & 2020-02-29 & 58908.05 & 1.6 & 5 & $2.5^{f}$ & $1.65_{-0.04}^{+0.04}$ & $1.00\pm0.20$ & $0.10^{f}$ & $2331_{-79}^{+19}$ & $0.12 \pm {0.01}$ & 220/187 \\
3200950102 & 2020-03-02 & 58910.32 & 1.6 & 4 & $2.5^{f}$ & $1.77_{-0.05}^{+0.05}$ & $1.00\pm0.40$ & $0.10^{f}$ & $2282_{-134}^{+19}$ & $0.09 \pm {0.00}$ & 153/146 \\
\hline

\end{tabular}
\begin{tablenotes}
\item[a] Errors quoted with 68\% confidence where a single value is given
\item[f] parameter frozen      
\end{tablenotes}
\end{threeparttable}
\end{table*}
\begin{table*}
\small
  \caption[]{Details of \textit{AstroSat} observations chosen for analysis are given in the first 7 columns. Spectral parameters fitted with model \\
  		\textit{TBabs (thcomp $\otimes$ diskbb)} are presented with 90\% confidence.}
     \label{tab:pheno_fit_astro}
\begin{threeparttable}      
\begin{tabular}{c@{\hspace{2pt}}c@{\hspace{2pt}}cc@{\hspace{2pt}}c@{\hspace{2pt}}c@{\hspace{2pt}}c@{\hspace{2pt}}c@{\hspace{2pt}}c@{\hspace{2pt}}ccc@{\hspace{2pt}}c@{\hspace{2pt}}c}
\hline
\smallskip
ObsId  &  Date & MJD & \multicolumn{2}{c}{SXT} & \multicolumn{2}{c}{LAXPC} & $n_{H}$ & $\Gamma$ & $f_{cov}^{a}$ & $T_{in}$ & N$_{diskbb}$ & $F_{0.5-10\;keV}$ & $\chi^{2}/dof$ \\
 & & & Exp. time & Rate & Exp. time & Rate & & & & & & & \\
 & & & \scriptsize{(ks)} & \scriptsize{\cps} & \scriptsize{(ks)} & \scriptsize{\cps} & \scriptsize{$(\times 10^{20}$ cm$^{-2})$} & & & \scriptsize{(keV)} & & \scriptsize{$(\times\;10^{-10}$ erg cm$^{-2}$ s$^{-1}$)} & \\
\hline
3290 & 2019-11-08 & 58795.09 & 5.6 & 34 & 6.1 & 154 & $2.6_{-0.5}^{+0.5}$ & $2.32_{-0.09}^{+0.08}$ & $0.04_{-0.01}^{+0.01}$ & $0.63_{-0.006}^{+0.006}$ & $2729_{-112}^{+19}$ & $73.60 \pm {0.10}$ & 486/392 \\
3306 & 2019-11-15 & 58802.07 & 8.0 & 33 & 14.7 & 92 & $2.5_{-0.5}^{+0.5}$ & $2.33_{-0.09}^{+0.09}$ & $0.04_{-0.01}^{+0.01}$ & $0.58_{-0.000}^{+0.005}$ & $2018_{-81}^{+19}$ & $39.80 \pm {0.10}$ & 508/397 \\
3328 & 2019-11-21 & 58808.28 & 6.0 & 38 & 11.4 & 96 & $2.5^{f}$ & $2.58_{-0.09}^{+0.09}$ & $0.17_{-0.02}^{+0.02}$ & $0.53_{-0.005}^{+0.005}$ & $2054_{-80}^{+19}$ & $29.50 \pm {0.10}$ & 478/393 \\
\hline
\end{tabular}
\begin{tablenotes}
\item[f] parameter frozen
\end{tablenotes}
\end{threeparttable}    
\end{table*}

\begin{table*}\small
  \caption[]{Details of \textit{Swift-XRT} observations chosen for analysis are given in the first 5 columns. Spectral parameters fitted with model \\
  		\textit{TBabs (thcomp $\otimes$ diskbb)} are presented with 90\% confidence}.
     \label{tab:pheno_fit_swift}
\begin{threeparttable}      
\begin{tabular}{c@{\hspace{2pt}}cc@{\hspace{2pt}}c@{\hspace{2pt}}cc@{\hspace{2pt}}c@{\hspace{2pt}}ccc@{\hspace{2pt}}cc}
\hline
\smallskip
ObsId & Date & MJD & Exp. time & Rate & $n_{H}$ & $\Gamma$ & $f_{cov}^{b}$ & $T_{in}$ & N$_{diskbb}$ & $F_{0.5-10\;keV}$ & $\chi^{2}/dof$ \\
 & & & (ks) & \cps & $(\times 10^{20}$ cm$^{-2})$ & & & (keV) & & $(\times\;10^{-10}$ erg cm$^{-2}$ s$^{-1})$ & \\
\hline
12172002 & 2019-11-08 & 58795.96 & 0.6 & 7.7 & $2.5^{f}$ & - & - & $0.61_{-0.030}^{+0.030}$ & $2889_{-495}^{+19}$ & $68.00 \pm {1.00}$ & 371/390 \\
12172003 & 2019-11-09 & 58796.74 & 1.4 & 7.2 & $2.5^{f}$ & $2.60_{-0.30}^{+0.50}$ & $0.20_{-0.20}^{+0.20}$ & $0.59_{-0.020}^{+0.020}$ & $3004_{-364}^{+19}$ & $69.90 \pm {0.30}$ & 324/321 \\
12172012 & 2019-11-19 & 58806.31 & 1.1 & 8.1 & $2.5^{f}$ & $2.60_{-0.20}^{+0.30}$ & $0.30\pm0.20$ & $0.48_{-0.020}^{+0.040}$ & $3929_{-953}^{+19}$ & $41.00 \pm {1.00}$ & 197/179 \\
12172014 & 2019-11-21 & 58808.89 & 1.7 & 5.9 & $2.5^{f}$ & $3.20_{-0.30}^{+0.50}$ & $0.30\pm0.10$ & $0.50_{-0.020}^{+0.050}$ & $3507_{-1065}^{+19}$ & $39.00 \pm {1.00}$ & 188/181 \\
12172015 & 2019-11-22 & 58809.29 & 1.4 & 8.0 & $2.5^{f}$ & $3.20_{-0.50}^{+0.90}$ & $0.30\pm0.20$ & $0.52_{-0.020}^{+0.020}$ & $3420_{-574}^{+19}$ & $39.00 \pm {1.00}$ & 232/190 \\
12172032 & 2019-12-12 & 58829.54 & 2.1 & 7.7 & $2.5^{f}$ & $3.00_{-1.00}^{+2.00}$ & $0.30_{-0.20}^{+0.20}$ & $0.43_{-0.030}^{+0.020}$ & $3465_{-531}^{+19}$ & $18.90 \pm {0.20}$ & 271/275 \\
12172066 & 2020-01-18 & 58866.61 & 0.7 & 11.1 & $2.5^{f}$ & $2.00_{-0.10}^{+0.10}$ & $0.21_{-0.04}^{+0.05}$ & $0.16_{-0.009}^{+0.009}$ & $36113_{-9292}^{+19}$ & $2.70 \pm {0.20}$ & 231/263 \\
12172077 & 2020-01-31 & 58879.22 & 1.7 & 1.1 & $2.5^{f}$ & $1.67_{-0.10}^{+0.05}$ & $0.60_{-0.20}^{+0.30}$ & $0.10^{f}$ & $11174_{-5139}^{+19}$ & $0.38 \pm {0.01}$ & 215/212 \\
\hline
\end{tabular}
\begin{tablenotes}
\item[b] Errors quoted with 68\% confidence where a single value is given
\item[f] parameter frozen
\end{tablenotes}
\end{threeparttable}       
\end{table*}

\subsection{\textit{AstroSat}}

\textit{AstroSat} observes sources simultaneously in soft and hard X-rays using \textit{Soft X-ray Telescope (SXT)} \citep{Singh2017} and \textit{Large Area X-ray Proportional Counter (LAXPC)} \citep{Yadav2016,Antia2017}. \textit{AstroSat} carried out ToO observation of MAXI J0637-430 in three different intervals on 8, 15 and 21 November 2019. The Level-1 data of \textit{SXT} and \textit{LAXPC} are obtained from the Indian Space Science Data Centre (ISSDC) archive \footnote{\tiny{\url{http://astrobrowse.issdc.gov.in/astro_archive/archive}}}. The details of the \textit{AstroSat} observations are given in Table \ref{tab:pheno_fit_astro}. \textit{SXT} Level-1 data obtained in the Photon Counting (PC) mode was processed through the latest pipeline (\textit{AS1SXTLevel2-1.4b}) to obtain Level-2 clean event data. \texttt{XSELECT V2.4g} was used to generate the image, lightcurve and spectra (see \cite{Sreehari2019,Baby2020} for details). Following the instructions in the \textit{SXT} user manual for data analysis, a source region with radius between 13 and 16 arcmin can be chosen for  extraction of spectral products\footnote{\tiny{\url{https://www.tifr.res.in/~astrosat_sxt/dataanalysis.html}}}. Since the count rate was higher than 40 \cps, which is the pile-up limit for \textit{SXT}, we first considered an annular region with an outer radius of 14 arcmin with a varying inner radius such that the count rate reduces below 40 \cps. Based on this criterion, we excluded a central circular region with radius 8 arcmin, 7 arcmin and 5 arcmin for observations on 8, 15 and 21 November respectively, to generate the lightcurve and spectra. \textit{SXT} image from 21 November 2019 is shown in Fig. \ref{fig:sxt_img}. Response and background files provided by the team were used. To correct for offset, \texttt{arf} file was generated using \texttt{sxtARFmodule} provided separately\footnote{\tiny{\url{https://www.tifr.res.in/~astrosat_sxt/dataanalysis.html}}}. \textit{SXT} data were grouped by 30 counts per bin for better statistics.

\textit{LAXPC} data was obtained in the Event Analysis (EA) mode for all the observations. Of the three \textit{LAXPC} detectors, the spectrum for \textit{LAXPC20} was generated using \texttt{laxpcsoftv3.3} version of \textit{LAXPC} software\footnote{\tiny{\url{https://www.tifr.res.in/~astrosat_laxpc/LaxpcSoft.html}}} \citep{Antia2017}. Only events detected in the top layer were selected along with a filter to reject multiple detections of a single event \citep{Agrawal2018,Baby2020}. Background was generated using the background model closest in time to the observations. The background contributed to more than 50\% of the observed flux beyond 25 keV. Hence, we performed the spectral analysis for \textit{LAXPC} in the $4-25$ keV energy band. The exposure times for both instruments are given in Table \ref{tab:pheno_fit_astro}. To generate PDS, lightcurves were obtained in the $3-12$ keV band with 5 ms binsize for comparison with \textit{NICER} PDS.

\subsection{\textit{Swift-XRT}}

As \textit{NICER} did not observe the source between 15 November 2019 and 25 November 2019 (see Fig. \ref{fig:maxi_fig}), we chose four observations of \textit{Swift-XRT} in this period in the Windowed Timing (WT) mode. Details of the observations are given in Table \ref{tab:pheno_fit_swift}. An additional four observations were chosen corresponding to different epochs during the outburst, to check for consistency in spectral parameters with other instruments. We present a subset of the observations already analysed in \cite{Tetarenko2021}. The $0.5-10$ keV spectra were generated using the standard online tools provided by the UK \textit{Swift} Science Data Centre \citep{Evans2009}. Count rate exceeded the pile-up limit of 100 \cps in the WT mode during the soft state. The online tools inherently adjust for this effect and hence all the observations were analysed using the online tools as recommended by the team (private communication with Kim Page). Slight variation is seen in the results obtained by \cite{Tetarenko2021} when compared with our results. The photon index is lower and the disc temperatures are slightly higher for all the observations in the soft state presented in \cite{Tetarenko2021}. Manual selection of source region was performed there, which probably affected the results due to severe pile-up issues in the WT mode observations of this source. We present a sample case in Appendix \ref{sec:append} where comparison of spectral parameters with variation in source region is discussed. The spectra were grouped to have a minimum of 5 counts per bin.

One-day averaged \textit{MAXI} lightcurve in the $2-20$ keV band is shown for reference in top panel of Fig. \ref{fig:maxi_fig}. \textit{AstroSat} and \textit{Swift-XRT} observations considered are plotted as green and red dashed vertical lines respectively in Fig. \ref{fig:maxi_fig}. Spectral fits are performed in \texttt{XSPEC v 12.11.1} \citep{Arnaud1996}.

\section{Analysis and Modelling}
  \label{sec:anal}
\subsection{Spectral Analysis}
  \label{sec:spec_anal}
The \textit{NICER} spectra are modelled in the $0.5-10$ keV range. Initially the source was reported to be in the soft state. Therefore, we model the spectra using an absorbed multicolour disc blackbody component alone, the \textit{diskbb} model in \texttt{XSPEC} \citep{Mitsuda1984,Makishima1986}. \textit{TBabs} \citep{Wilms2000} is used to account for the interstellar absorption with abundance set to \texttt{wilms} with \texttt{vern} cross-section \citep{Verner1996}. The spectral modelling resulted in poor fits with $\chi^{2}_{red}=1135/578$. A very steep powerlaw was required to account for the residuals above 5 keV giving a much better value of $\chi^{2}_{red}=415/576$.  A systematic error of 1.5\% was added during the spectral fitting \citep{Jana2021}. Although instrumental residuals are observed below 1 keV, they do not affect the fits as they are within the systematics limit of the instrument i.e., a few percent at 1 keV (private communication with Craig Markwardt). However, the powerlaw component seems to contribute significantly in the lower energy range, which is unlikely in the soft state. We then convolve the disc with the thermal Comptonisation model \textit{thcomp} \citep{Zdziarski2020}, a more accurate version of \textit{nthcomp}. The parameters are similar to those in \textit{nthcomp} with the addition of a Comptonisation fraction term ($f_{cov}$). Since the electron plasma temperature ($kT_e$) cannot be constrained, owing to the cut-off energy possibly lying beyond the \textit{NICER} energy range considered, we freeze it to 100 keV. The final model used is \textit{TBabs (thcomp $\otimes$ diskbb)}. The $n_H$ parameter ranges from $\sim$ $1-3 \times 10^{20}$ cm$^{-2}$ with most of the values lying between 2 and 3 $\times 10^{20}$ cm$^{-2}$. We, therefore, freeze $n_H$ to an average value of $2.5 \times 10^{20}$ cm$^{-2}$ in cases where it could not be constrained towards the end of the outburst.

We model the broadband spectrum ($0.5-25.0$ keV) from \textit{AstroSat} with \textit{TBabs (thcomp $\otimes$ diskbb)} model as described above. Response is modified for both \textit{SXT} and \textit{LAXPC} spectra using the \texttt{gain} command \citep{Singh2017,Antia2021}. A systematic error of 2.5\% is applied to the combined fit \citep{Mukerjee2020}. Positive residuals are observed in the $5-7$ keV range in the energy spectrum obtained on 21 November 2019 (ObsId 3328). To test for evidence of reflection, we replace \textit{thcomp} with the relativistic reflection model \textit{relxillCp} \citep{Garcia2014,Dauser2014} which includes a physical Comptonisation continuum. Both models fit the spectrum well with  $\chi^2_{red}$ of 476/390  and 478/393 with and without the inclusion of reflection respectively. However, individual parameters in the model could not be constrained. Such features are not observed in other available observations with \textit{AstroSat}. Addition of either a reflection component from the returning radiation or a single-temperature blackbody component, associated with the boundary of the black hole, was required for spectral fits in observations close to this date using \textit{NuSTAR} data in \cite{Lazar2021}. A blackbody component (\textit{bbodyrad}) is included to check for the requirement of such an additional component. This results in poorer fits with $\chi^{2}_{red}$ $>2$. Moreover, a few parameters are pegged at their respective limits. Therefore, we fit all the spectra with the model \textit{TBabs (thcomp $\otimes$ diskbb)}. The \textit{Swift-XRT} spectra in the $0.5-10$ keV range can also be fitted using the same model. However, the $n_H$ value could not be constrained and was fixed at $2.5 \times 10^{20}$ cm$^{-2}$. 
The fit parameters using all the instruments are presented in Tables \ref{tab:pheno_fit_nicer}, \ref{tab:pheno_fit_astro} \& \ref{tab:pheno_fit_swift}. All errors are quoted at 90\% confidence unless stated otherwise.   
\subsection{Temporal Analysis}

Lightcurves in the $0.3-12$ keV energy band for \textit{NICER} were generated with a binsize of 5 ms to study the evolution of the temporal properties of the source.  PDS was obtained in the rms space using \texttt{powspec} command from \texttt{XRONOS v6.0}. Lightcurves were divided into intervals of 16384 bins and Poisson noise subtracted PDS was obtained for each interval. Averaging all the PDS in the frequency range $0.01-100$ Hz gives the final PDS. Geometric rebinning by a factor of 1.05 is performed. All the \textit{NICER} observations upto MJD 58858 could be modelled by a powerlaw.  Band-limited noise was seen during the transition from soft to hard state and LHS in the \textit{NICER} data, which was modelled with zero centred Lorentzian (Fig. \ref{fig:timing_fit}). The rms values are computed over the frequency range of 0.01 to 100 Hz \citep{Remillard2006}.

The $3-12$ keV lightcurves with the same binsize of 5 ms were generated with \textit{LAXPC20} for better comparison with \textit{NICER} results. PDS were obtained in the rms space following the procedure detailed above. The first two observations with \textit{AstroSat} could be modelled by a powerlaw. However, broad Lorentzian components were required to model the PDS of \textit{LAXPC20} observation of 21 November 2019 as discussed in Section \ref{sec:temp_prop}.
\section{Results}
\subsection{Outburst Profile and HID}
 \label{sec:hid}
%
 \begin{figure}
	\centering
	\includegraphics[scale=0.46]{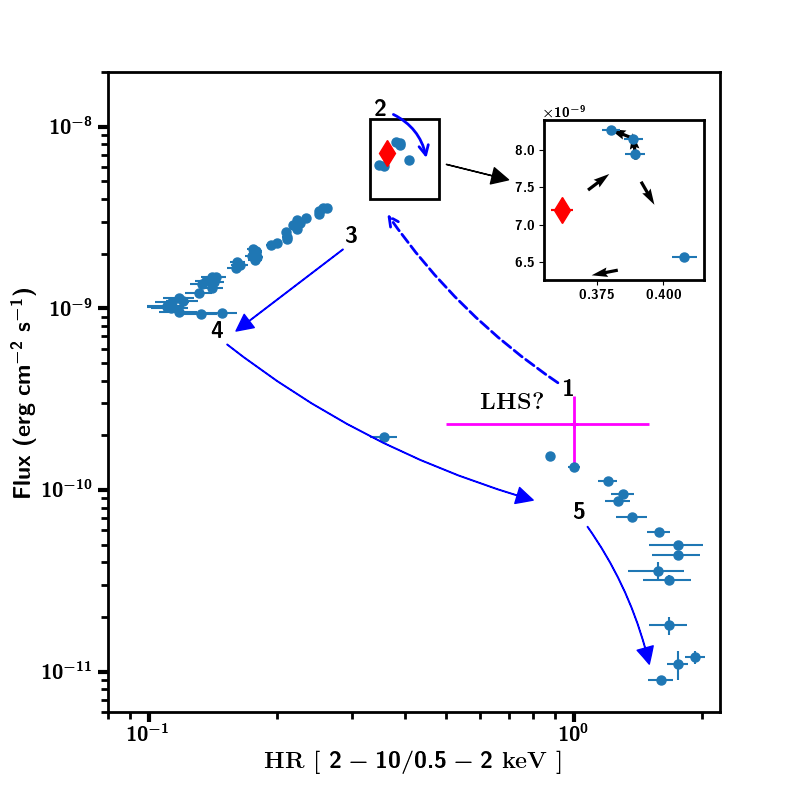}
	\caption{Flux in the $0.5-10$ keV band is plotted against the hardness ratio obtained using \textit{NICER}. The hardness ratio is defined as ratio of flux in the energy range 2--10 keV and 0.5--2 keV. The evolution of the HID is shown using the arrows from the points 1 through 5. The magenta data point is flux in the $0.5-10$ keV band obtained from the \textit{MAXI} count rate of the source on 2 November 2019 using the Crab spectrum (see text for details). The magenta data point and the dashed arrow represent the probable location of the source at the beginning of the outburst where it could have been in the LHS and its subsequent evolution. A zoomed in version of the rectangular patch is also shown as inset to highlight the small circular loop the source follows within the rectangle at point 2. The red diamond denotes the first observation obtained using the \textit{NICER} data.}
	\label{fig:hid}
\end{figure}
\begin{figure*}
	\centering
	\includegraphics[scale=0.35,angle=-90]{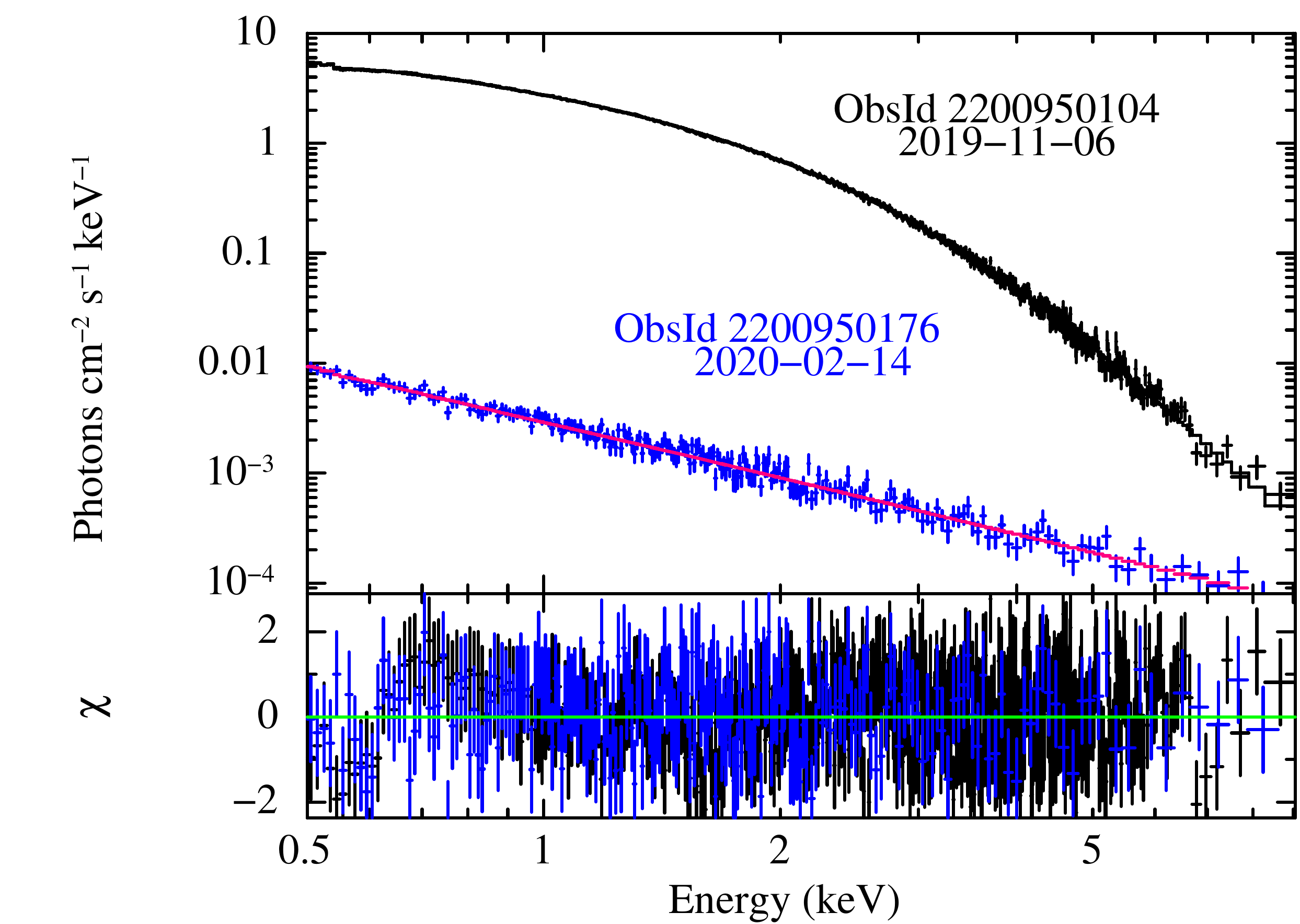}\includegraphics[scale=0.35,angle=-90]{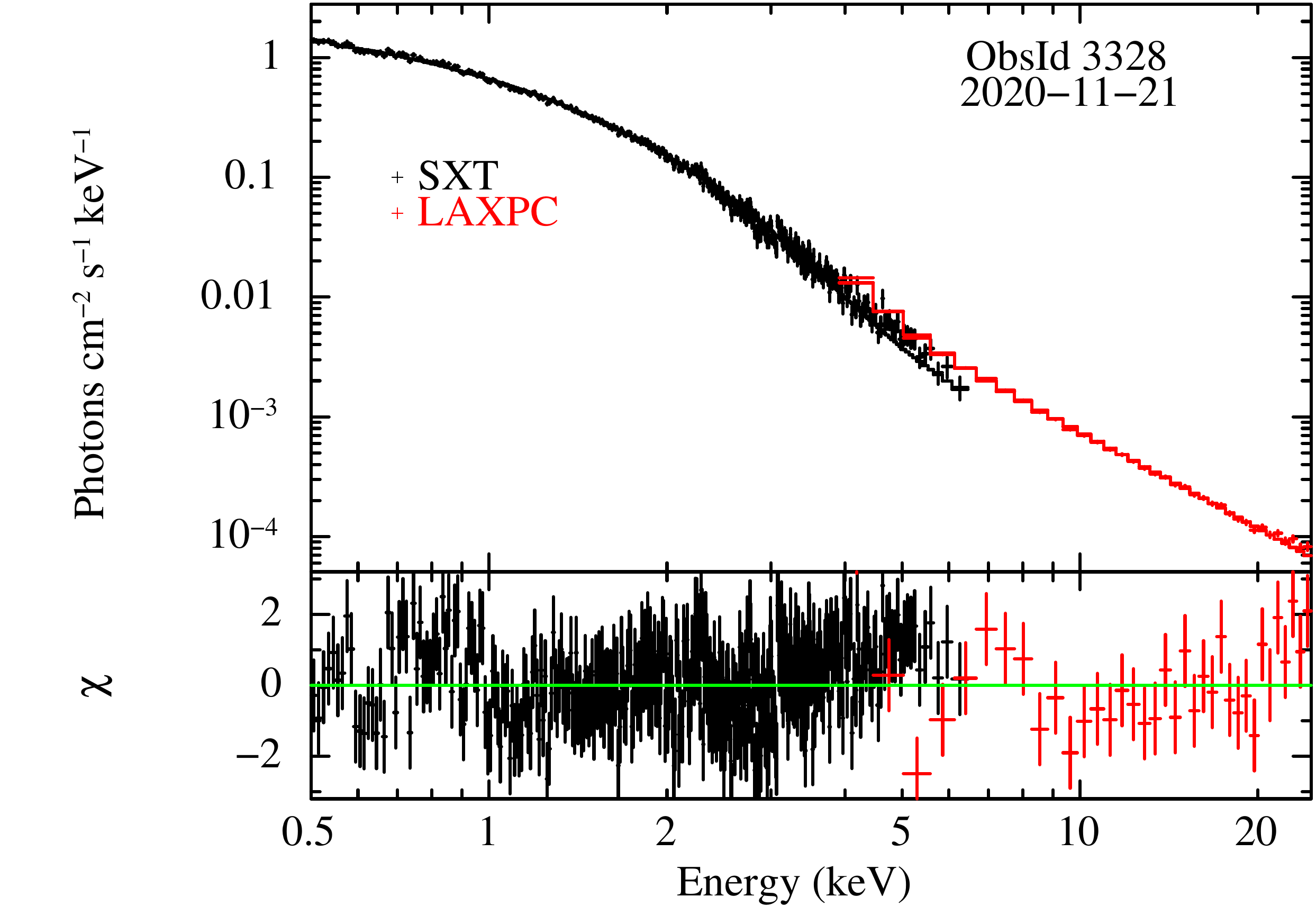}
	\caption{Spectra of \textit{NICER} and \textit{AstroSat} fit with the model \textit{TBabs(thcomp $\otimes$ diskbb)} are shown. The left panel shows the unfolded \textit{NICER} spectra from 6 November 2019 (black) and 14 February 2020 (blue) to show the difference between spectral states. The unfolded spectra of 21 November 2019 from \textit{AstroSat} is plotted in the right panel. The black data points represent \textit{SXT} and red represent \textit{LAXPC}. The residuals are plotted in the bottom panels.}
	\label{fig:spec_fit}
\end{figure*}
The top panel of Fig. \ref{fig:maxi_fig} shows the \textit{MAXI} lightcurve for the source in the $2-20$ keV energy band. \textit{NICER} lightcurve, for all the GTIs combined and binned by 100 seconds for clarity, is shown in the middle panel. The corresponding hardness ratios are plotted in the bottom panel.  As can be seen from the top panel of Fig. \ref{fig:maxi_fig}, the source brightened to $\sim$ 90 mCrab within 2 days of its detection. Thereafter, a gradual decay is seen for the next $\sim$ 25 days. The source remained faint for the rest of the outburst, eventually reaching quiescence towards the end of April (not shown in the plot). HR is defined as the ratio of counts in the $2-10$ keV band to the counts in $0.5-2$ keV band. HR shows a slight rise towards the beginning of the outburst and then dips lower than 0.1. Purely based on the variation in HR, this would suggest that the source could have been in an intermediate state from MJD 58790 to MJD 58801. The source then remained in soft state till MJD 58858. An evident rise in HR is again seen after this date till MJD 58880 where the HR reaches 0.4 and remains constant, corresponding to intermediate state and LHS, respectively. The green and red dashed vertical lines plotted in Fig. \ref{fig:maxi_fig} denote the \textit{AstroSat} and \textit{Swift-XRT} observations respectively. Details of \textit{NICER, AstroSat} and \textit{Swift-XRT} observations are given in Tables \ref{tab:pheno_fit_nicer}, \ref{tab:pheno_fit_astro} and \ref{tab:pheno_fit_swift}. 

 We also plot the HID using \textit{NICER} in Fig. \ref{fig:hid}. The unabsorbed model flux obtained from fitting \textit{NICER} spectra in the $0.5-10$ keV range is plotted along with the hardness ratio i.e. the ratio of flux in the $2-10$ keV to $0.5-2$ keV energy bands. The source traces a `c' shaped profile in the HID. However, \textit{NICER} data is available only from 3 November 2019. We, therefore, use \textit{MAXI} data available before this date to obtain a complete picture of evolution of the HID. Crab spectrum from \textit{MAXI} is accumulated within the duration of the observations used here (MJD 58790$-$58910) using \textit{MAXI} on-demand process\footnote{\tiny{\url{http://maxi.riken.jp/mxondem/}}}. The correlation factor between the flux in the $2-10$ keV range and count rate is calculated. This number is then used to correlate the  count rate and flux in $2-10$ keV in MAXI J0637-430 on 2 November 2019 in Crab units. Total flux is obtained by extrapolating the response to lower energies. Similar procedure is used to obtain flux in the $0.5-2$ keV band. This point is taken only as a reference as the calibration difference between the instruments is not considered. Therefore, the evolution from points 1 to 2 is marked as dotted line denoting the possible path of evolution. Solid arrows from the points 2 through 5 show the evolution of the source as seen using \textit{NICER}. Interestingly, the source seems to follow a loop within the rectangle at point 2 before moving to the bright intermediate state. This is emphasized in the inset of Fig. \ref{fig:hid}. The arrows show the direction of evolution. The HR decreases from 0.5 to 0.1 (point 3 to 4) where the source is in the HSS. Then the source moves to the LHS where a sudden increase in HR is seen (0.1 to 0.8) and flux is reduced by an order ($10^{-8}$ to $10^{-9}$ erg cm$^{-2}$ s$^{-1}$) (point 4 to 5). To further confirm the scheme of classification, we look at the spectro-temporal properties of the source in detail using all the instruments in the next section. 

\subsection{Spectral properties}
%

   \begin{figure}
   \centering
   \includegraphics[scale=0.5]{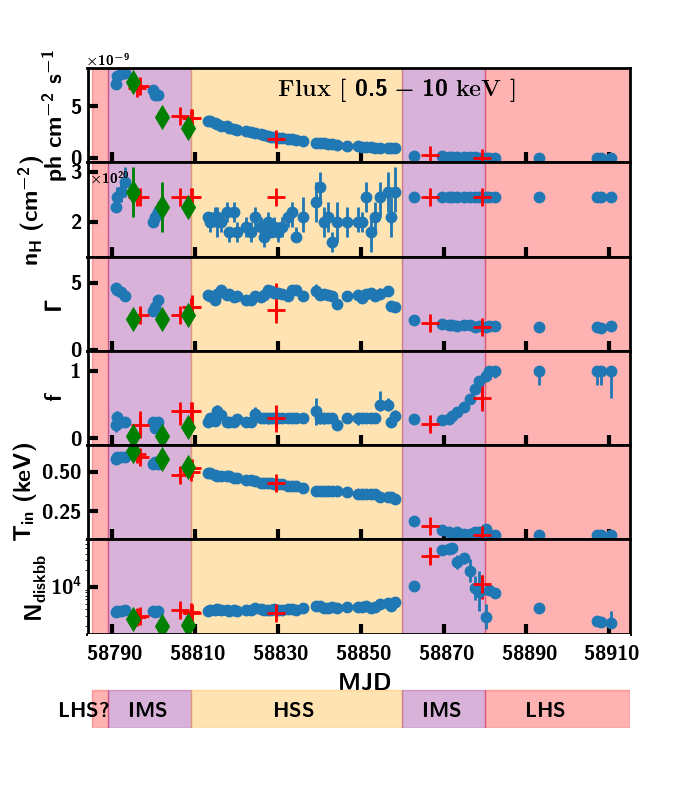}
      \caption{Variation in parameters obtained from fits with \textit{TBabs(thcomp $\otimes$ diskbb)} are shown.  The filled blue circles, green diamonds and red crosses represent \textit{NICER}, \textit{AstroSat}  and \textit{Swift-XRT} observations. Flux in the $0.5-10$ keV is plotted in the first panel. Variation in $n_{H}\;(\times 10^{20} cm^{-2})$, photon index ($\Gamma$), scattering fraction ($f$), disc temperature $T_{in}$ (keV) and diskbb normalization ($N_{diskbb}$) are plotted in panels 2 to 6 from the top. The red, purple and yellow patches correspond to LHS, IMS and HSS respectively (see text for details).}
         \label{fig:par_var}
   \end{figure}
   \begin{figure}
   \centering
   \includegraphics[scale=0.5]{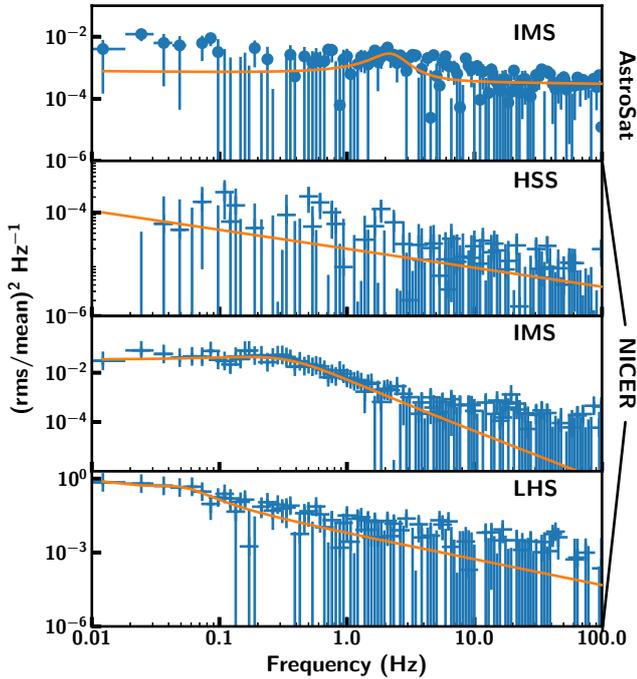}
      \caption{PDS for the source in HSS, IMS and LHS is shown above. The top panel shows the \textit{LAXPC20} PDS from 21 November 2019 where the source was in the IMS with filled circles. The second, third and fourth panels show the PDS from \textit{NICER} observations on 29 November 2019, 21 January and 14 February 2020 respectively. The solid line shows the best fit model consisting of only powerlaw for HSS and a combination of broad Lorentzians and powerlaw for IMS and LHS.}
         \label{fig:timing_fit}
   \end{figure}

Energy spectra are modelled with \textit{TBabs(thcomp $\otimes$ diskbb)} as detailed in Section \ref{sec:anal}. The results of the fit are presented in Tables \ref{tab:pheno_fit_nicer}, \ref{tab:pheno_fit_astro} and \ref{tab:pheno_fit_swift}. The $n_{H}$ parameter is left free and varies from $1.7-2.8$ $\times 10^{20}\; cm^{-2}$ except for cases where it is fixed at $2.5 \times 10^{20}$ cm$^{-2}$ as mentioned previously. Fig. \ref{fig:spec_fit} shows the unfolded spectrum for 21 November 2019 from \textit{AstroSat} in the right panel and 6 November 2019 and 14 February 2020 from \textit{NICER} in the left panel. As the outburst progresses, the contribution of the disc component reduces and the spectrum becomes harder. The Compton scattering fraction ($f_{cov}$) increases significantly after 25 January 2020. The energy spectra from 2 February 2020 are a simple powerlaw distribution. Therefore, it is difficult to constrain the parameters of the \textit{diskbb} component. We fix the temperature at 0.1 keV to perform the fits in this case as shown in Table \ref{tab:pheno_fit_nicer}. An example spectrum from 14 February 2020 is shown in the left panel of Fig. \ref{fig:spec_fit}.

The evolution in parameters is shown in Fig. \ref{fig:par_var}. Disc temperature and photon index decrease from 0.58 to $\sim$ 0.1 keV and $\sim$ 4.6 to $\sim$ 1.7, respectively as the source evolves in the decay phase (Table \ref{tab:pheno_fit_nicer} and Fig. \ref{fig:par_var}). The total flux is obtained in the range $0.5-10$ keV for all the observations using the \textit{cflux} command. The parameter $f_{cov}$ acts as a marker for the rise in prominence of the Comptonisation component seen in the fourth panel of Fig \ref{fig:par_var}. However, lack of high energy spectra limits the accurate estimation of this parameter. In many cases, a precise estimation of $f_{cov}$ is not possible during the soft states, and therefore, we define the upper limit to be 0.3 and quote the error at 68\% confidence (see Tables \ref{tab:pheno_fit_nicer}, \ref{tab:pheno_fit_swift}). The existence of broadband spectra upto 25 keV allows us to obtain better estimates of $f_{cov}$ during \textit{AstroSat} observations. It can be seen from Table \ref{tab:pheno_fit_astro} that $f_{cov}$ is $<$ 0.1 on 8 and 15 November 2019. A slight increase in $f_{cov}$ and low values of $\Gamma$ (2.3-2.6) are seen during \textit{AstroSat} observations, suggesting that the source could have been in an intermediate state. Reflection features are not observed in the spectrum. In the absence of \textit{NICER} data, we also evaluate \textit{Swift-XRT} data to confirm this transition and find that a decrease in $\Gamma$ is noted here as well. However, better constraints on the parameters cannot be obtained due to a lack of high energy data. Based on the evolution of spectral parameters, we now divide the outburst into 3 states, i.e., LHS, Intermediate State (IMS) and HSS. The red, purple and yellow patches in Fig. \ref{fig:par_var} denote these states. Based on only the \textit{NICER} lightcurve and HR (Fig. \ref{fig:maxi_fig}), the source seems to have moved to the soft state at the peak of the outburst. However, the addition of \textit{AstroSat} and \textit{Swift-XRT} data show a clear dip in $\Gamma$ accompanied by a variation in the value of $f_{cov}$. It seems to imply that the source did not transition to HSS and was still in an intermediate state. Hence, we extend the intermediate state to at least MJD 58809. The source remains in the soft state from MJD 58813 to 58880 and then reaches the canonical hard state, where $f_{cov}$ reaches the upper limit of 1 and a corresponding decrease in $\Gamma$ is seen.
\subsection{Temporal Properties}
\label{sec:temp_prop}
The top panel of Fig. \ref{fig:timing_fit} shows the PDS obtained from \textit{AstroSat} data on 21 November 2019 when the source was in an intermediate state. An increase in rms values is observed using \textit{LAXPC20} PDS from 11\% to 20\% from 8 November to 21 November 2019 respectively. Additional broad Lorentzian component was also required to model the PDS obtained on 21 November 2019 with centroid frequency at $\sim$ 2 Hz. The PDS also shows a small bump at 0.02 Hz. We attempt to model it using a Lorentzian component. However, as the change in $\chi^{2}_{red}$ is very small (124/118 to 118/115), we model the PDS using a powerlaw component and a single broad Lorentzian. The PDS for all the observations from 3 November to 15 December 2019 using \textit{NICER} exhibit weak red noise which are modelled with a powerlaw. A sample PDS is shown in the second panel of Fig. \ref{fig:timing_fit} which corresponds to 26 November 2019. The rms remains less than 5\% for these observations, typical of a BHB in the HSS \citep{Remillard2006,Motta2018,Baby2020}. Band-limited noise (BLN), characteristic of a harder state, is observed in the PDS for the observations  with \textit{NICER} after 21 January 2020, modelled by zero centred Lorentzians as shown in third and fourth panels of Fig. \ref{fig:timing_fit}. An increase in rms from $18.7\pm0.5$\% on 21 January 2020 to $24\pm2$\% on 30 January 2020 is seen, which is close to the expected rms from BHBs in LHS \citep{Remillard2006,Nandi2012,Belloni2014}. The rms continues to increase till 3 February 2020 and reaches upto $\sim$ 30\%. Although the source remains in a canonical hard state in the following observations, the flux from the source decreases further, limiting the estimation of rms. The PDS become largely dominated by noise and can be modelled by a powerlaw component.

\section{Discussion}

This paper analyses and presents data analysis results from three instruments - \textit{NICER, AstroSat} and \textit{Swift-XRT}. We study the outburst profile and perform spectral and timing analysis of the source in detail through different phases of the outburst. We also attempt to comment on  the physical nature of the source using physical models.

\subsection{Probing the nature of the compact object : Obtaining limits on mass and distance}

The X-ray transient MAXI J0637-430 was discovered in November 2019 and was extensively observed during its outburst. Spectral and temporal studies of the source have been performed providing some details as to the nature of the source \citep{Tetarenko2021,Jana2021}. Based on the spectral parameters, it is proposed to be a black hole candidate. We attempt to obtain a lower limit on distance using the normalization of the \textit{diskbb} model ($N_{diskbb}$) resulting from spectral fits to an observation during the disc dominant state. It is defined as $N_{diskbb} = (R_{in}/D_{10})^2\; cos\; \theta$, where $R_{in}$ is the apparent inner disc radius, $D_{10}$ is the distance to the source in units of 10 kpc and $\theta$ is the inclination angle. We use this relation to obtain a lower limit on the distance considering inclination in the range $5-80^{\circ}$ \citep{Motta2018} and mass as 3 $M_{\sun}$ as the compact object is more likely to be a black hole. The colour to effective temperature ratio ($f_{col}$) is considered to be 1.7 \citep{Shimura1995}.  Assuming a non-rotating BH and $N_{diskbb}$  $\sim$ 3800, the distance to the source is greater than 2 kpc.

\begin{figure}
   \centering
   \includegraphics[scale=0.5]{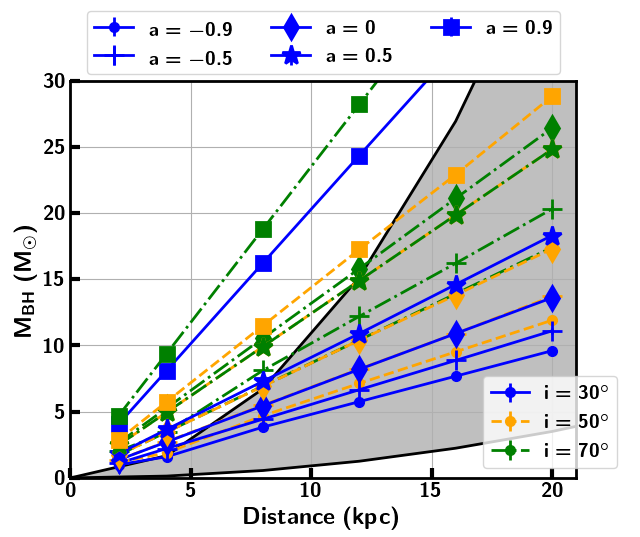}
      \caption{Mass is plotted as a function of distance and inclination derived using \textit{kerrbb} model. Inclination and distance were varied in steps of 10$^{\circ}$ and multiples of 2 respectively. Solid, dashed and dotted lines denote inclination angles 30$^{\circ}$, 50$^{\circ}$ and 70$^{\circ}$ respectively. Spins of -0.9,-0.5,0,0.5 and 0.9 are considered. The shaded region denotes the most probable distance and mass estimates obtained using the soft to hard transition luminosity.}
         \label{fig:distmassinc}
   \end{figure}

Since the source was predominantly in the disc dominant state initially, we also attempt to model the spectrum using relativistic disc model \textit{kerrbb} \citep{Li2005} instead of \textit{diskbb} to constrain the mass and spin of the BH with the model combination of \textit{TBabs (thcomp $\otimes$ kerrbb)}. \textit{kerrbb} model can be used when the contribution of the disc to the total flux is more than 70\%. Although this criterion is satisfied by the observations before MJD 58862, we choose the observation with the maximum flux on 6 November 2019. Due to many unknown parameters, viz., distance, mass, inclination angle and spin, it is challenging to obtain constraints on the physical parameters using this model. As the lower limit of distance is obtained as 2 kpc, we systematically increase the distance in the broad range 2 to 20 kpc in multiples of 2 and spin from -0.9 to 0.9 (-0.9, -0.5, 0 , 0.5, 0.9) to obtain the mass of the compact object at a fixed spectral hardening factor of 1.7 \citep{Shimura1995}. The variation of mass with distance and spin is presented in Fig. \ref{fig:distmassinc}. The inclination angle is frozen in steps of 10$^{\circ}$ between 30$^{\circ}$ and 70$^{\circ}$. The curves obtained with inclination angles 30$^{\circ}$,  50$^{\circ}$ and 70$^{\circ}$ are shown in Fig. \ref{fig:distmassinc} for clarity. For inclination angles less than 30$^{\circ}$, mass is found to be less than 3 $M_{\sun}$.  Since the aim is to estimate the upper limit on mass with  this exercise, we limit ourselves to inclination angles greater than 30$^{\circ}$. We also plot mass as a function of distance using the luminosity of the source as it transits from  soft to hard state. This transition generally occurs at $1-4$ \% of $L_{Edd}$ \citep{Maccarone2003,Dunn2010,Tetarenko2016,Motlagh2019}. The shaded region in Fig. \ref{fig:distmassinc}, gives the range of mass obtained following this assumption. From the overlap of the shaded region and the distance-mass curves obtained, mass is constrained within $3-19$ $M_{\sun}$ for distance $<$ 15 kpc with retrograde spin. Mass of the BH is $>20$ $M_{\sun}$ when a non-rotating BH or one with prograde rotation is considered. The orbital period of the system is estimated to be $2-4$ hrs \citep{Tetarenko2021}. Such low orbital period Galactic systems are unlikely to contain a BH of mass $>20$ $M_{\sun}$ \citep{Motta2018,Jonker2021}. Therefore, we propose that MAXI J0637-430 consists of a BH with mass in range $3-19$ $M_{\sun}$ with retrograde rotation at a distance $<$ 15 kpc. 

   \begin{figure}
   \centering
   \includegraphics[scale=0.5]{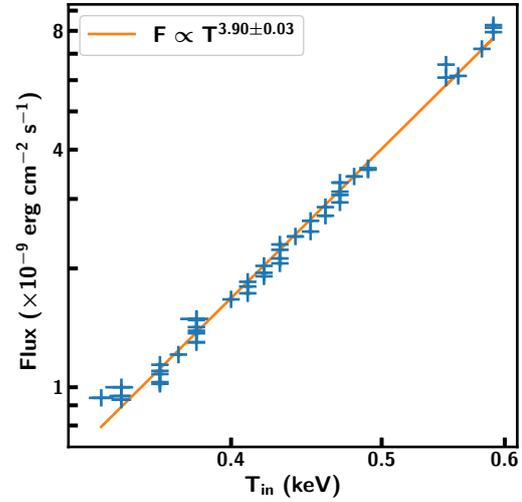}
      \caption{Flux in the $0.5-10$ keV band is plotted against the disc temperature. The solid line shows the best fit obtained. Flux is proportional nearly equal to fourth power of temperature which signifies that the disc conforms to the standard picture.}
         \label{fig:temp_vs_flux}
   \end{figure}
\subsection{Tracking the evolution of the source - deviations from the standard picture}
The source seems to track out a distinctive `c' in the HID (see Fig. \ref{fig:hid}) with a circular knot at the top instead of the `q'-shaped HID expected during a canonical outburst. A similar profile was seen for the BHB 4U 1630-472 \citep{Baby2020}. A typical outburst remains in the hard state initially, which forms the right stem of the `q' in the HID. In contrast, MAXI J0637-430 and 4U 1630-472 seem to undergo quick transitions from hard to soft state. In fact, `mini'-outbursts of 4U 1630-472 were characterized by a lack of hard states at the beginning of the outburst \citep{Abe2005,Capitanio2015}. However, it was found that 4U 1630-472 undergoes a quick transition from hard to soft state (within a few hours) and was probably missed in the previous outbursts \citep{Baby2020}. It is difficult to localize the starting point of the outburst in the HID as \textit{NICER} data before 3 November 2019 is not available. The \textit{MAXI} data point (magenta cross) plotted in Fig. \ref{fig:hid} acts as a reference to the beginning of the outburst as detailed in Section \ref{sec:hid}. It suggests that the source was in the hard state for a very short duration at the beginning of the outburst and a similar quick transition from hard to disc dominant state could have occurred within the time span of a few hours. Additionally, the existence of an intermediate state can be inferred at the beginning of the outburst based on the spectral characteristics, although a clear distinction from the soft state is difficult to observe in the absence of broadband spectra.  Earlier studies classify the source as being in the soft state for most of the duration of the outburst  \citep{Tetarenko2021,Jana2021,Lazar2021}. This could be due to the absence of broadband data or the unavailability of continuous observations. For instance, lower values of $\Gamma$ and a slight increase in rms are also reported for observations with \textit{Swift+NuSTAR} from MJD 58801 to 58812 \citep{Lazar2021}. The results obtained in our analysis, taken in conjunction with those presented in \cite{Lazar2021}, suggest that the source was not in a canonical soft state at the beginning of the outburst. A transition to a harder state during this part of the outburst was also postulated for 4U 1630-472 \citep[see Fig 4 in][]{Baby2020}. 

Due to the availability of broadband spectra from \textit{AstroSat}, where $\Gamma$ and $f_{cov}$ are better constrained, it can be stated that the source undergoes transition to an intermediate state lasting for a few days before going to the soft state. Random variation in spectral parameters in this duration (MJD 58790 to MJD 58809) is also seen (Tables \ref{tab:pheno_fit_nicer}, \ref{tab:pheno_fit_astro}, \ref{tab:pheno_fit_swift} and Fig. \ref{fig:par_var}). Although, presence of a reflection component cannot be ruled out based on fits with \textit{relxillCp} model as discussed in Section \ref{sec:spec_anal}, clear positive residuals in the $5-7$ keV range are seen only in the spectrum from MJD 58808.28. Therefore, we restrict our discussion to the evolution of spectral parameters in the context of a two-component model. The rapid fluctuations in spectral parameters (within hours or a few days) can be associated with the changes in the disc during a predominantly soft state where the assumption of a standard Shakura-Sunyaev disc \citep{Shakura1973} breaks down. Such deviations are expected at the beginning and end of disc-dominated states or intermediate states which can be explained by a moving disc or change in colour correction factor ($f_{col}$) as disc fraction decreases \citep{Dunn2011}, corresponding to a change in degree of ionization in the disc. To check for deviations from the standard picture during the soft state, we plot the disc temperature vs. flux in Fig. \ref{fig:temp_vs_flux}. The solid line shows the best fit relation between the two parameters ($T_{in}$ and $F$) as
$F \propto T^{3.90\pm0.03}$. This is clearly close to the standard relation of $F \propto T^{4}$ expected of SS discs. It suggests that the source evolved to the canonical HSS quickly, which is unlikely if it is driven only by the disc accretion. Such a scenario in 4U 1630-472 is explained by \cite{Capitanio2015} by invoking an external perturbation that triggers a rise in disc temperature independent of the disc accretion rate. The results obtained so far suggest that the physical mechanism underlying the origin of outbursts for MAXI J0637-430 and 4U 1630-472 could be similar. However, further inferences cannot be made unless extensive broadband observations of the sources are performed for the subsequent outbursts and recurrence times are closely monitored. 

The spectral and temporal properties of the source are consistent with those of a black hole binary. Assuming a lower mass limit of 3 $M_{\sun}$ for the compact object, we obtain the lower limit of distance as 2 kpc. The source was also found to undergo transitions from IMS $\rightarrow$ HSS $\rightarrow$ IMS $\rightarrow$ LHS as also reported by \cite{Tetarenko2021} and \cite{Jana2021}. Additionally, we could extend the duration of the initial intermediate state as compared to the one proposed by \cite{Jana2021} with the addition of broadband \textit{AstroSat} data. Based on the spectral fits and transitional luminosity relation, the mass of the compact object is $<$ 19 $M_{\sun}$ at distance $<$ 15 kpc with retrograde spin. The `c' shaped HID and the quick transition from hard to soft state at the beginning of the outburst suggest that the standard picture of varying contributions from thermal accretion disc and Comptonisation component due to disc truncation is insufficient to explain the complex evolution of the spectra. Further, studies with future observations of the source will aid in understanding its many peculiarities.
%

\section*{Acknowledgements}
      We thank the reviewer for his/her constructive feedback and comments which greatly improved the quality of this paper. This publication uses the data obtained through High Energy Astrophysics Science Archive Research Center (HEASARC) online service, provided by the NASA/Goddard Space Flight Center. This research has made use of data obtained from \textit{AstroSat} mission of Indian Space Research Organisation (ISRO), archived at the Indian Space Science Data Centre (ISSDC). We thank the \textit{SXT} and \textit{LAXPC} POC at TIFR, Mumbai for verifying and releasing the data. We also thank the respective instrument teams for assistance on the data reduction and analysis. This work made use of data supplied by UK Swift Science Data Centre. This research has made use of \textit{MAXI} data provided by \textit{RIKEN, JAXA and MAXI team}. The authors acknowledge the financial support of ISRO under \textit{AstroSat} archival Data utilization program Sanction order No. DS-2B-13013(2)/13/2019-Sec.2. AN thanks GH, SAG, DD, PDMSA and Director, URSC for encouragement and continuous support to carry out this research.
\\
\\

\textit{Facilities : NICER, AstroSat, Neil Gehrels Swift-XRT}

\section*{Data Availability}
The data for \textit{NICER} and \textit{Swift-XRT} underlying this article are publicly available in the High Energy Astrophysics Science Archive Research Center (HEASARC) at \url{https://heasarc.gsfc.nasa.gov/db-perl/W3Browse/w3browse.pl}. \textit{AstroSat} data is available at \textit{AstroSat}-ISSDC website (\url{http://astrobrowse.issdc.gov.in/astro_archive/archive}).

\bibliographystyle{mnras} 
\bibliography{ref_maxij0637}
%

\appendix
\section{Data reduction of \textit{Swift-XRT} : Effects of pileup}
  \label{sec:append}
\begin{figure}
	\centering
	\includegraphics[scale=0.23]{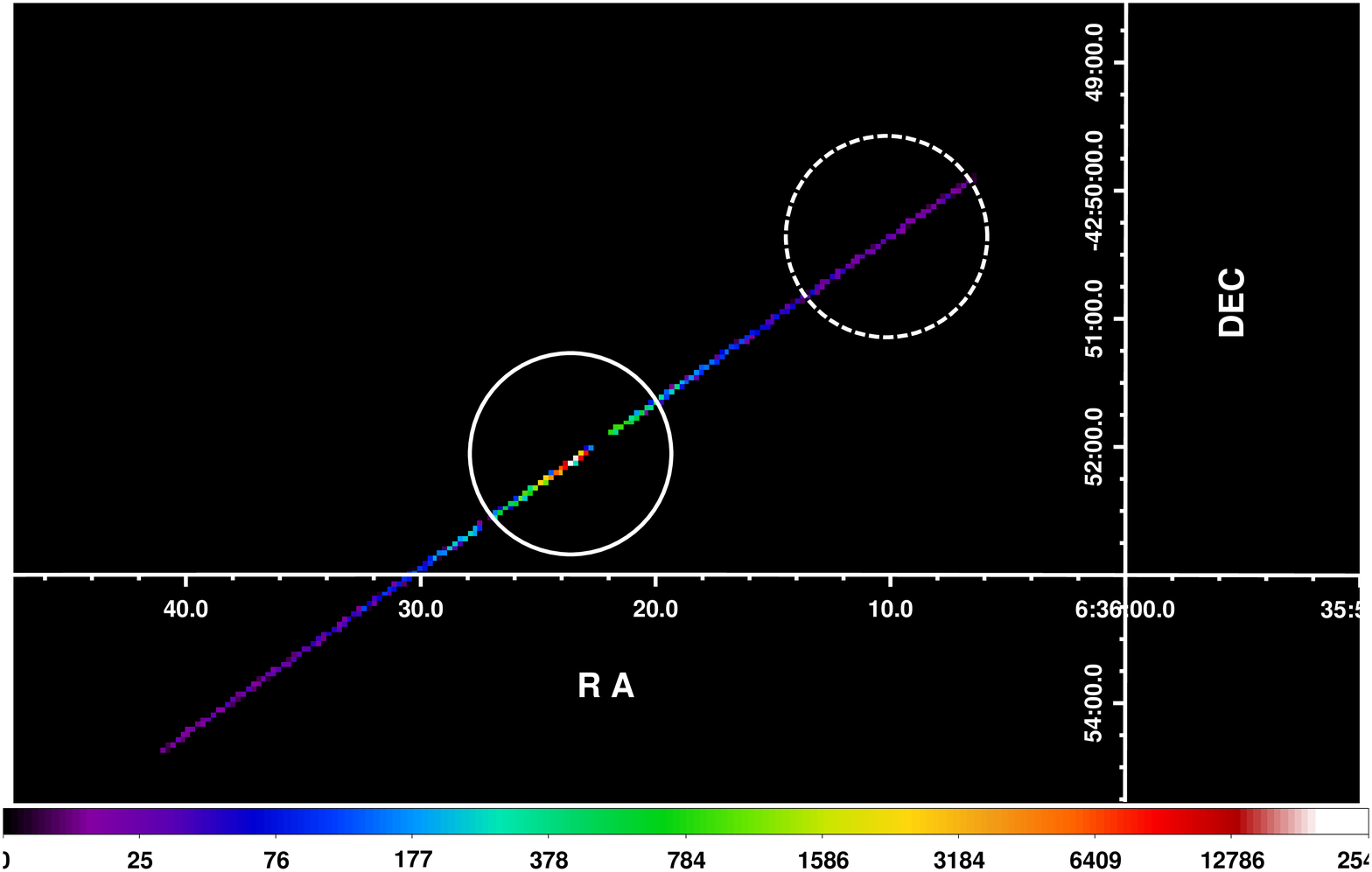}
	\caption{\textit{Swift-XRT} image of the source from ObsId 12172002 obtained on 8 November 2016. The solid circle and the dashed circle are source and background regions considered, each of 20 pixel radius.}
	\label{fig:appenda}
\end{figure}
\begin{figure}
	\centering
	\begin{subfigure}[a]{0.55\textwidth}
		\includegraphics[scale=0.35, angle=-90]{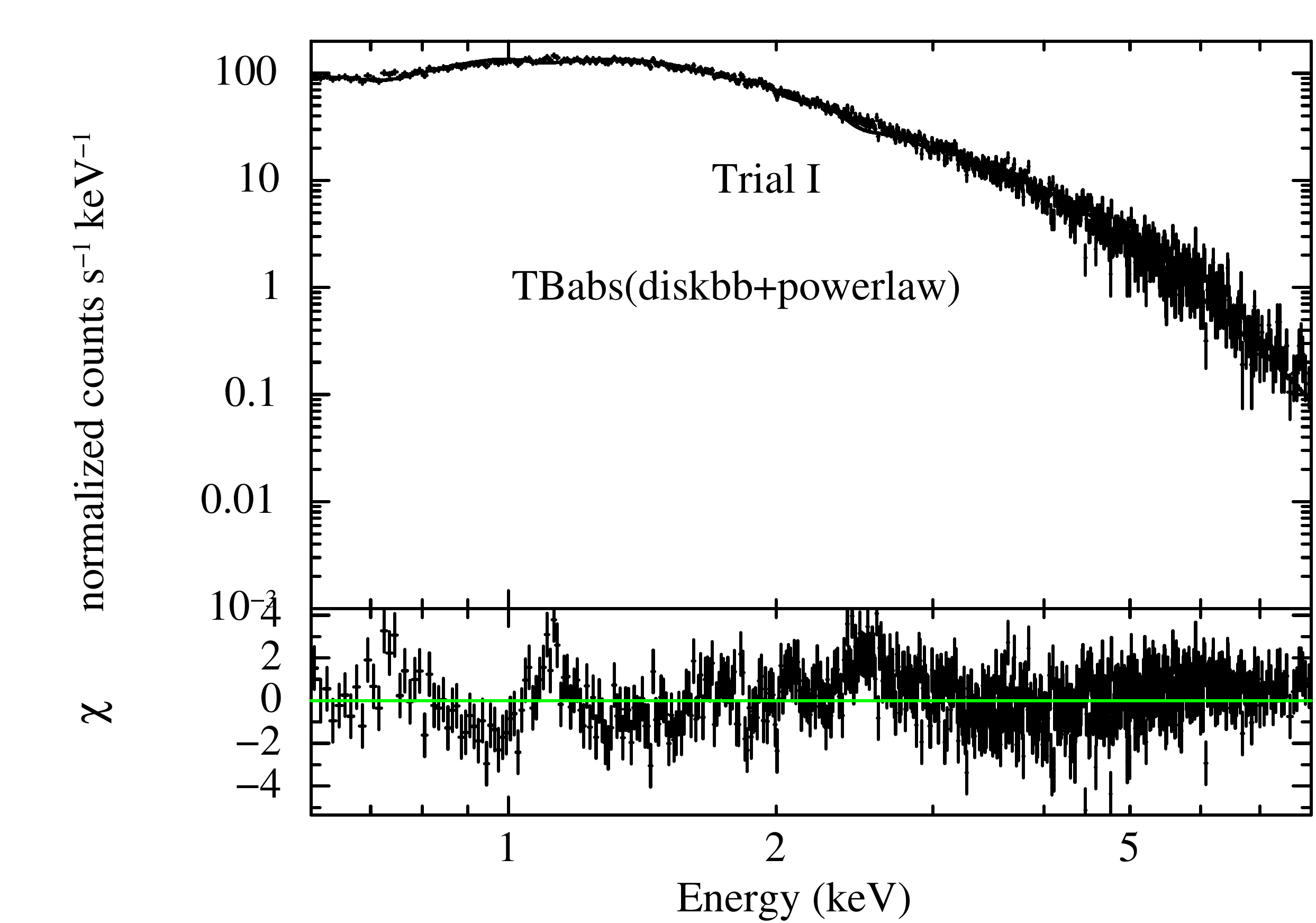}
		\label{fig:appendba}
	\end{subfigure}
	\begin{subfigure}[b]{0.55\textwidth}
		\includegraphics[scale=0.35, angle=-90]{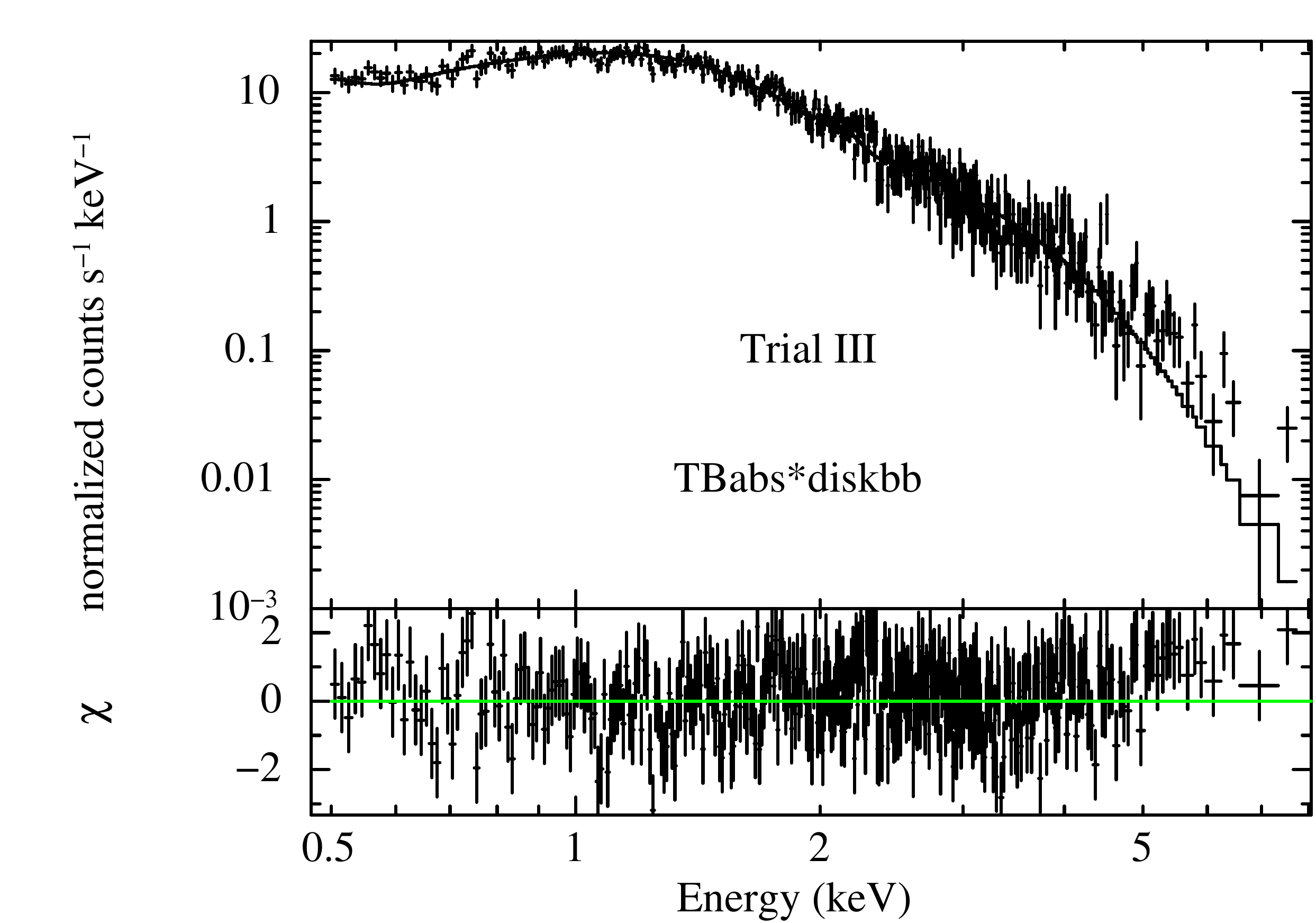}
		\label{fig:appendbb}
	\end{subfigure}
	\caption{The top two spectra show the model fits to ObsId 12172002 where circular source extraction region is considered (top panel) and an annular source extraction region is taken (bottom panel). The spectra are clearly affected by pileup as the requirement of a powerlaw component is introduced to fit the spectrum in the top panel.}
	\label{fig:appendb}
\end{figure}

MAXI J0637-430 was particularly bright in the $0.5-10$ keV band from the beginning of the outburst to the end of the soft state thereby making it susceptible to pile-up issues in CCD based detector instruments like \textit{Swift-XRT} and \textit{SXT} onboard \textit{AstroSat}. The effects of pile-up in \textit{SXT} are accounted for by an annular region for source extraction in a way that the count rate falls below 40 \cps as detailed in Section \ref{sec:anal}. In this section, we describe the effects of pile-up in \textit{Swift-XRT} spectra. Recently, \cite{Tetarenko2021} performed spectral analysis of \textit{Swift-XRT} data and have reported harder values of $\Gamma$ compared to those quoted in the present work. Although pile-up issues in the PC mode data are addressed in \cite{Tetarenko2021}, effect of pile-up in the WT mode was not taken into consideration. We report that the difference in spectral parameters arises as WT mode data is affected by pile-up when offline data reduction is performed using \texttt{xrtpipeline}. To demonstrate the change in spectral parameters, we choose one observation from \textit{Swift-XRT} observations as a sample and obtain spectra by using different source selection regions.

Most of the observations for this source are available in Windowed Timing (WT) mode and a few in Photon Counting (PC) mode. Data is considered to be piled-up if the count rate exceeds 100 \cps in WT mode and 0.5 \cps in PC mode. In all the observations of the source in soft state, the count rate exceeds 100 \cps and reaches almost 250 \cps during the peak of the outburst. We therefore consider one of the brightest observations with the source close to the peak of the outburst on 8 November 2019 (ObsId 12172002). We then obtain the final spectra for three cases which we name as Trials I, II and III. 

In Trial I, we follow \cite{Tetarenko2021} and consider a circular source region with radius of 20 pixels centred around its coordinates. A corresponding background region is chosen towards the edge with the same radius. The source and background regions are shown in Fig. \ref{fig:appenda}. The response file is generated using \texttt{xrtmkarf} module and the spectrum is grouped to have a minimum of 5 counts per energy bin. We fit the spectrum obtained with the phenomenological model with \textit{TBabs (diskbb+powerlaw)}. We also freeze the $n_H$ value to 4.39 $\times 10^{20}$ cm$^{-2}$ as given in \cite{Tetarenko2021}. The disc temperature and $\Gamma$ are obtained as 0.78 $\pm$ 0.05 keV and 1.5 $\pm$ 0.2 respectively. This is close to the value obtained for the same observation in \cite{Tetarenko2021}. However, as the background subtracted count rate is $>$ 200 \cps, we select an annular source region excluding a circular region of radius 5 pixels in the centre in Trial II and 10s pixel in Trial III, maintaining the net radius of source region at $r=r_2 - r_1 = 20$, where $r_1$ is the inner radius and $r_2$ is the outer radius. Corresponding response files are generated each time using \texttt{xrtmkarf} module. The spectrum can be fit using only an absorbed multicolour disc blackbody model. The spectrum fits well in both cases with $\chi^{2}_{red}$ of 510/419 and 397/361 in Trial II and Trial III respectively. We also included an additional powerlaw component for comparison. However, the fits did not improve and the component was discarded based on the results of \texttt{ftest}. The temperature of the disc reduces to 0.70 in Trial II and 0.67 in Trial III.  Comparison of the two model fitted spectra from Trial I and Trial III are shown in Fig. \ref{fig:appendb}.  It is clear from both panels that a significant difference in the source spectrum is observed as a result of pile-up effects.


\bsp	
\label{lastpage}
\end{document}